\documentclass[journal]{IEEEtran}

\usepackage{xcolor,soul,framed} 

\colorlet{shadecolor}{yellow}
\usepackage[pdftex]{graphicx}
\graphicspath{{../pdf/}{../jpeg/}}
\DeclareGraphicsExtensions{.pdf,.jpeg,.png}

\usepackage[cmex10]{amsmath}
\usepackage[update,prepend]{epstopdf}
\usepackage[noadjust]{cite}
\usepackage[latin1]{inputenc}
\usepackage{tikz}
\usepackage{cite}
\usetikzlibrary{angles, quotes}
\usepackage{bbm} 
\usepackage{pdfpages}
\usepackage{tabulary}
\usepackage{multirow}
\usepackage{comment}
\usepackage{dsfont}
\usetikzlibrary{calc}
\usepackage{ amssymb }
\usepackage[mathscr]{eucal}
\usepackage{bm}
\usepackage{comment}
\usepackage{adjustbox}
\usepackage{tikz}
\usepackage{pgfplots}
\usepackage[linesnumbered,lined,boxed,commentsnumbered]{algorithm2e}
\usepackage{amssymb}
\usepackage{textcomp}
\usepackage{graphicx}
\usepackage{graphics}
\usepackage{epsfig}
\usepackage{epstopdf}
\usepackage{float}
\usepackage{color}
\usepackage[cmex10]{amsmath}
\usepackage{latexsym,amsfonts}
\usepackage{amsthm}
\usepackage{url}
\usepackage{longtable}
\usepackage[figuresright]{rotating}
\usepackage{listings}
\usepackage{etoolbox}
\usepackage[latin1]{inputenc}
\usepackage{algorithmic}
\usepackage{array}
\usepackage{commath}
\usepackage{xcolor}
\usepackage{caption}
\usepackage{multicol}
\usepackage{matlab-prettifier}
\usepackage{booktabs}
\usepackage{tabularx}

\begin{document}

    \title{Maritime Communications: A Survey on Enabling Technologies, Opportunities, and Challenges}
  \author{Fahad S. Alqurashi, {\em Student Member, IEEE}, Abderrahmen Trichili, {\em Member, IEEE}, Nasir Saeed, {\em Senior Member, IEEE}, Boon S. Ooi, {\em Senior Member, IEEE} and Mohamed-Slim Alouini, {\em Fellow, IEEE}

\thanks{F. S. Alqurashi, A. Trichili, B. S. Ooi, and M.-S Alouini are with the Computer, Electrical and Mathematical Sciences $\&$ Engineering in King Abdullah University of Science and Technology, Thuwal, Makkah Province, Saudi Arabia. Email: \{fahad.alqurashi,abderrahmen.trichili,boon.ooi,slim.alouini\}@
kaust.edu.sa.\newline 
N. Saeed is with the, Department of Electrical and Communications Engineering, United Arab Emirates University (UAEU), Al Ain, UAE. Email: mr.nasir.saeed@ieee.org.}
}

\maketitle

\begin{abstract}
Water covers 71\% of the Earth's surface, where the steady increase in oceanic activities has promoted the need for reliable maritime communication technologies. The existing maritime communication systems involve terrestrial, aerial, and space networks. This paper presents a holistic overview of the different forms of maritime communications and provides the latest advances in various marine technologies. The paper first introduces the different techniques used for maritime communications over the radio frequency (RF) and optical bands. Then, we present the channel models for RF and optical bands, modulation and coding schemes, coverage and capacity, and radio resource management in maritime communications. After that, the paper presents some emerging use cases of maritime networks, such as the Internet of Ships and the ship-to-underwater Internet of things. Finally, we highlight a few exciting open challenges and identify a set of future research directions for maritime communication, including bringing broadband connectivity to the deep sea, using terahertz and visible light signals for on-board applications, and data-driven modeling for radio and optical marine propagation.
\end{abstract}



%
\IEEEpeerreviewmaketitle

\section{Introduction} 
The recent progress in terrestrial communication technologies has unlocked unprecedented data rates, enabling various new applications. With the upcoming sixth generation (6G) era, the capacity growth is expected to improve by 10 to 100 times \cite{DavidIEEEVTechMag18, DangNatElec20}. Despite the recent advancements in wireless communications on land, offering reliable and high-speed data rates for marine communication remains challenging. Maritime communication is crucial due to the dramatic increase in oceanic activities, including naval shipping and logistics, offshore oil exploration, wind farming, fishing, and tourism, among others. Maritime communication is equally needed for Internet of things (IoT) applications, such as environmental monitoring and climate change control. Nevertheless, fourth generation (4G) and fifth generation (5G) are limited in maritime environments as base stations cannot be installed far offshore and in oceans, restricting their use to onshore scenarios only.
For these reasons, satellites are a crucial pillar in maritime communications. Satellites are essential in providing connectivity in unconnected oceans at the expense of high proprietary terminal and operation costs, in addition to the limited available bandwidth \cite{xu2017quality, jiang2015possible}. Recently, aerial wireless solutions such as unmanned aerial vehicles (UAVs) and helikites are being introduced to maritime communication applications \cite{XianlingTCOM20,BlueComPlus}. 
Presently, most of the technologies for ship-to-ship and ship-to-shore communication use medium frequency (MF), high frequency (HF), very high frequency (VHF), and ultra-high frequency (UHF) bands. Although these bands can ensure relatively long propagation distances, they support only basic applications such as half-duplex voice calling, text messaging, and the automatic identification system (AIS). Therefore, improving the quality of life on board when sailing for long distances requires more advanced communication systems beyond those offered by satellites or maritime radio.
Radio maritime communication is subject to various impairments, including rain causing significant signal scattering and sea waves causing vibrations affecting antenna height and orientation. Water surface reflections also cause a number of rays to reflect and interfere with each other. \newline
Besides the radio frequency (RF) band, optical wireless communication-based solutions in the infrared band, known as free space optics (FSO), also provide connectivity for maritime networks. Due to the collimated nature of laser beams, optical-based maritime communication systems are not affected by water surface reflections. However, FSO signals are affected by sea waves and weather conditions. Sea waves lead to pointing errors between FSO terminals, while weather conditions such as fog create scatterings of optical signals. Turbulence caused by the random variations of the refractive index of the atmospheric channel is also a major concern for FSO links causing scintillation and random light beam movements at the detector plane. 
There are also hybrid solutions proposed for maritime communication that install FSO on top of the RF infrastructure, which aim to provide more robust communication to FSO and RF propagation effects while having higher data rates. Various models have been proposed in the literature for RF and optical wave propagation through oceanic environments.\newline
\subsection{Related Surveys}
In recent years, there has been a growing interest in developing maritime communication networks (MCNs), and multiple surveys related to this topic have been published \cite{YangBook14,zolich2019survey,jo2019lte,guan2021magicnet,aslam2020internet, WeiIoT21, BalkeesICACCI15, WangAccess18}. These articles covered various aspects of maritime communication, such as the different network architectures, RF channel models, communication and networking of autonomous marine systems, and the IoT in maritime environments. For instance, a comprehensive survey on video transmission scheduling for wideband maritime communication is presented in \cite{YangBook14}. Then, Zolich \textit{et al.} reviewed the major advancements in autonomous maritime systems and applications and also provided an overview of maritime communication and networking technologies \cite{zolich2019survey}. Furthermore, the authors in \cite{jo2019lte} discussed the progress on a long-term evolution maritime (LTE-maritime) Korean project aiming to provide high data rates in orders of 10 Mbps with 100 km coverage. Authors of \cite{guan2021magicnet} briefly reviewed current maritime communication and networking projects and introduced the key technologies and applications of a novel maritime giant cellular network (Magicnet) architecture based on seaborne floating towers acting as base stations to provide wide coverage.
The authors of \cite{aslam2020internet} provided a comprehensive survey on the applications and challenges of maritime IoT technologies, also known as the Internet-of-Ships (IoS). Maritime communications and IoTs enabled by hybrid satellite-territorial networks were surveyed in \cite{WeiIoT21}. Moreover, there are also a few surveys on maritime communications focusing on RF channel models \cite{BalkeesICACCI15, WangAccess18}. A summary of the area of focus of related surveys is given in Table~\ref{Tab:SurveysComp}. 
\begin{table}[htbp]
\centering
\caption{Summary of related surveys.}
\begin{tabularx}{\linewidth}{|l|l|X|}
\hline
\multicolumn{1}{|c|}{{\textbf{Ref.}}}& \multicolumn{1}{c|}{{\textbf{Year}}} & \multicolumn{1}{c|}{{\textbf{Area of focus}}}   \\
\hline
\cite{YangBook14}&2014& Presents video transmission scheduling in maritime
wideband communication networks.\\
\hline
\cite{zolich2019survey}&2019& Surveys advancements in autonomous maritime systems with an overview of current and future communication technologies.\\                                           
\hline
\cite{jo2019lte} & 2019& Provides an overview of existing maritime communication systems and introduces LTE-maritime network, a project being conducted in South Korea, aiming to provide 100 km marine coverage. \\
\hline
\cite{guan2021magicnet} & 2019&  Presents the pros and cons of existing maritime communication technologies and proposes a maritime giant cellular network architecture (MagicNet).\\
\hline
\cite{aslam2020internet}& 2021& Provides an overview of the key elements and main characteristics of the IoS paradigm.\\                                                 
\hline
\cite{WeiIoT21}& 2021& Discusses hybrid satellite-terrestrial maritime networks.\\   
\hline
 \cite{BalkeesICACCI15,WangAccess18}& 2018 & Surveys different RF channel models for maritime communications.\\ 
 \hline 
 \end{tabularx}
\label{Tab:SurveysComp}   
\end{table}

\begin{figure*}[h]
    \centering
    \includegraphics[width=0.95\linewidth] {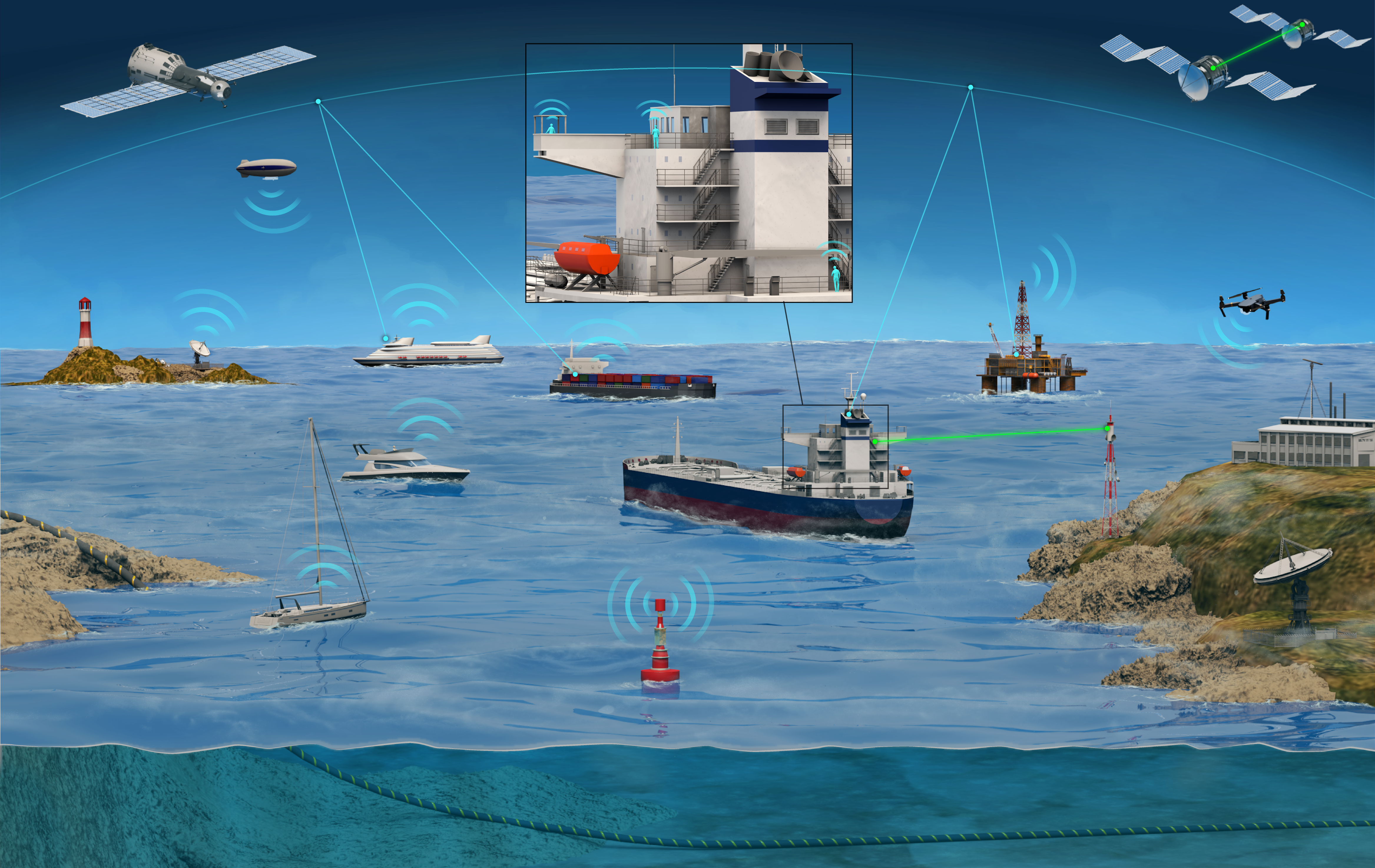}
    \caption {Maritime communication use cases.}
    \label{Fig:MaritimeCom}
\end{figure*}
\subsection{Contributions}
Motivated by the emerging research in maritime communication, this survey presents a holistic vision of marine communication and related technologies. We note that none of the previously presented references performs a complete survey of various aspects of maritime communication. Due to the prominent role of communication in marine life and industry, it is essential to study the different building blocks of maritime communication and the related emerging technologies. This article covers the latest advances in technologies connecting ships, ships and shore, and those sailing on-board far from shores. We discuss the building blocks of maritime communication, including radio resource management, coverage and capacity, and modulation and coding schemes. We additionally present a handful of emerging applications of maritime IoT. We particularly highlight the emerging progress on the maritime internet of things (IoT). Although underwater communication is out of the scope of the present survey, we cover the relationship between reliable maritime communication and IoUT. A few open issues related to maritime communication are highlighted. Moreover, future research directions are discussed, including using visible light communication (VLC) and terahertz (THz) band for on-board communication and inter-medium communication. Harnessing and data-driven modeling for future maritime channel modeling.

The paper is organized as follows:\newline
\noindent - In Section \ref{Sec:Overview}, we provide an overview of the various forms of maritime communication, namely RF, optical wireless, and hybrid RF/optical solutions for ship-to-ship, ship-to-shore, satellite-ship, and on-board communication.\newline 
\noindent - In Section \ref{Sec:BBlocks}, we presents the building blocks of maritime communication, covering the different propagation effects and channel models as well as modulation schemes and resource management.\newline
\noindent - In section Section \ref{Sec:IoSParadigm}, we introduce the internet of ships (IoS) and maritime IoT paradigms, and follow it by a discussion on IoUT.\newline
\noindent - In Section \ref{Sec:Challenges}, we discuss the challenges and open problems of maritime communication and identify future research directions. \newline
We conclude the paper with a few remarks on the need for reliable maritime communication in Section \ref{Sec:Conclusion}.
\section{Overview of Maritime Communications}
\label{Sec:Overview}
Ancient Greek ships used speaking trumpets that intensified and directed the human voice as a means of marine communication in the 5th century BC. For many centuries, homing pigeons and small boats were used to convey physical messages from ship to shore and ship to ship. Semaphore flag signaling became the principal means of maritime communication by the 18th century. Each flag represents a letter or signal. Light torches in the nighttime replaced flags. Until today, semaphore signaling is still recognized as a means of maritime communication. The development of the electromagnetic theory by Maxwell and the telephone invention by Marconi in the 19th century allowed for the wireless transfer of messages in the form of Morse codes. Using Morse codes over radio waves is also known as wireless telegraphy and was ensured by radio operators transferring and receiving messages at rates up to 200 words per minute. In the early 1900s, multiple naval warships were equipped with radiotelephones or ``voice radio''. The idea is to convert sound waves into radio at the transmitter using amplitude modulation and then convert the received radio signals back to sound waves at the destination using frequencies ranging from 2 to 23 MHz. In 1950, a VHF band was allocated for marine use. In 1962, the Commercial Telecommunications Satellite Act, \textit{a US federal statute}, was put into effect, allowing the launching of satellites into outer space for telecommunication purposes \cite{SatelliteAct1962}. This act supported the introduction of satellites in maritime communication. Many international organizations, including the International Association of Lighthouse Authorities, the International Telecommunication Union (ITU), and the International Maritime Organization, recognize the benefits of seamless data exchange for maritime communities. Nowadays, various platforms, including satellites, high-altitude platforms (HAPs), and UAVs, operating in different frequency bands, such as RF and optical, are used to provide maritime coverage, as seen in Fig.~\ref{Fig:MaritimeCom}. In the following, we will present the various forms of maritime communications and highlight the latest progress in each of these technologies.\newline
\begin{figure*}[h]
    \centering
    \includegraphics[width=1\linewidth] {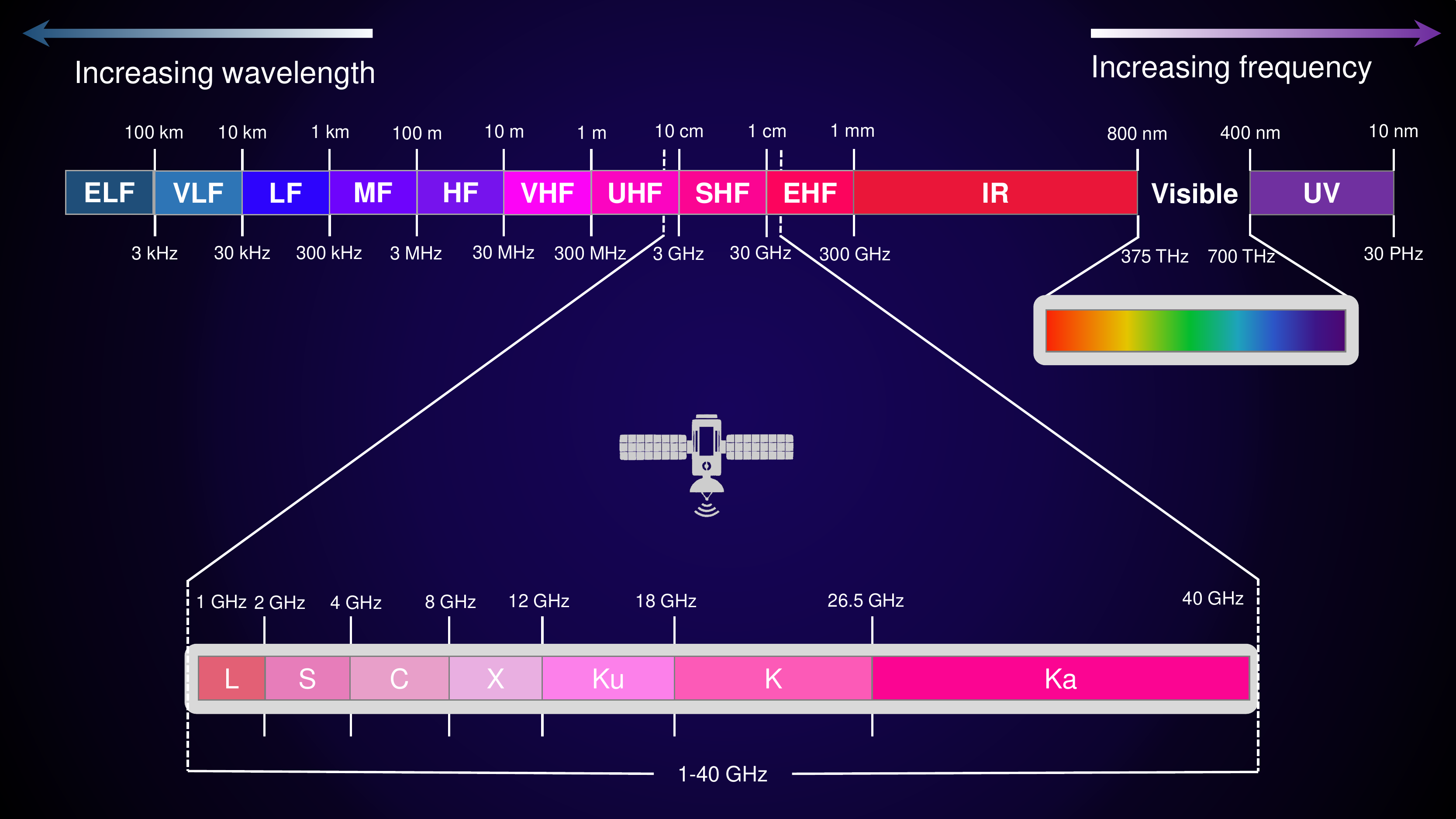}
    \caption {Electromagnetic spectrum in frequency and wavelength. ELF: extremely low frequency; VLF: very low frequency; LF: low frequency; MF: middle frequency; HF: High frequency; VHF: very high frequency; UHF: ultra high frequency: SHF: super high frequency; EHF: extremely high frequency; IR: infrared; UV: ultraviolet. Frequency bands in the region between 1-40 GHz are denoted by letters by IEEE.} 
    \label{fig:EMSpectrum}
\end{figure*}

\subsection{RF Technologies for Maritime Communications}
\label{Subsec:RFTech}
Maritime radio provides commercial and recreational communications uses and allows search and rescue assistance to ships in distress. An ITU-designated band for marine radio in the VHF band is the VHF maritime mobile band from 156 and 174 MHz. Marine VHF transceivers are installed in all large vessels and most seagoing crafts where a particular VHF frequency, known as Channel 16 (156.8 MHz), is designated as an international distress frequency. VHF terminal can be portable on a vessel or installed, allowing higher transmission power. Various VHF systems allow the following functionalities:
\begin{itemize}
\item Voice-only: which relies on the human voice for calling and communicating.
\item Digital selective calling (DSC): In addition to the voice calling functionality, the DSC allows the user to communicate with another vessel using a unique identifier known as the Maritime Mobile Service Identity (MMSI). The MMSI information is transmitted digitally, and once detected by a receiver, the operator of the receiving vessel will be alerted of the incoming call. DSC allows the automatic transfer of the caller's coordinates when sending a distress call if connected to a global positioning system (GPS).  
\item Automatic identification system (AIS): AIS allows the digital transfer of MMSI together with other information, including the vessel specification, real-time coordinates, speed, and course, to avoid collisions. AISs also transfer application-specific messages (ASM) to ships in ports and underway for safe navigation and boost maritime security. AIS operates as a mesh network, extending the communication ranges and enabling access to maritime traffic. 
\item Text messaging: It is also possible to send and receive text messages between VHF terminals using the  Radio Technical Commission for Maritime RTCM 12301.1 standard. 
\end{itemize}
According to the IMO, fitting an AIS transceiver is required in any cargo ship over 300 gross tonnage and all people transporting vessels. AIS transceivers use two ITU-designated VHF frequencies, known as marine band channel 87B at 161.975 MHz and channel 88B at 162.025 MHz. Each AIS transmits and receives over two channels to avoid interference. AIS transmission is based on a Gaussian minimum shift keying (GMSK) frequency modulation (FM) at a data rate of 9.6 kbps. Each time frame lasts 60 seconds and is divided into 2250 time slots, where each slot is 26.67 ms and contains 256 bits of data. AIS equipment uses self-organized time-division multiple access (SOTDMA) datalink schemes that are responsible for data transmission by using a reference time derived from GPS signaling to synchronize numerous data streams sent from many AIS transponders on a single band channel \cite{AISRadioScience}. 
VHF Data Exchange System (VDES) is another VHF-enabled technology, which is seen as \textit{the successor of AIS} offering the same functions of an AIS with the ability to connect to satellite using the same antenna and providing a higher data rate.
Radio maritime communication is also possible on the MF and HF bands. For example, navigation data (NAVDAT) is a safety and security maritime digital broadcasting system that operates in the MF/HF bands. In the MF/HF bands, radio waves are reflected at ionospheric layers, enabling longer propagation distances, reaching several hundreds of kilometers. NAVDAT is set to complement and possibly replace the direct printing navigational TEleX (NAVTEX) system operating in the MF band. NAVDAT also provides extended coverage compared to NAVTEX, enabling a maximum offshore range of 200 nautical miles ($\sim$ 370 km). In the following, we discuss several wireless access technologies in the RF band for maritime communications.\newline
\subsubsection{Standard Wireless Access Technologies}
\label{StdWirelessAccessNet}
\hfill\\
Efforts have been made to provide high-speed data connectivity using standard wireless access techniques beyond voice calls, text messaging, and safety information exchange.\newline
For instance, the TRI-media Telematic Oceanographic Network (TRITON) project was conducted to provide broadband internet offshore using a wireless mesh network of ships connected to terrestrial networks via an onshore station. TRITON is based on the IEEE 802.16 (Worldwide Interoperability for Microwave Access (WiMAX)) \cite{TRITON13}. Within the project Mare-Fi, aiming to provide WiFi broadband maritime communication for fishing ships and small vessels off-cost, a 1 Mbps data rate at a distance of 7 km was demonstrated \cite{MareFi14}. 
The MariComm, maritime broadband communication system was launched to target beyond 1 Mbps transfer rates at a 100 km distance from the shore \cite{MariComm15} in a multihop configuration. The multihop relay network was tested with 4 vessels acting as relays and reported a maximum data rate of 2 Mbps with a total distance of 70 km using LTE for the shore-to-ship link and wireless LAN (WLAN) for ship-to-ship communication.
BLUECOM+ is another project that was launched with the aim of providing broadband internet connectivity at large distances. Simulation results revealed that BLUECOM+ could provide 3 Mbps communication at 150 km from shore \cite{BlueComPlus}.
BLUECOM+ leverages the following to provide such connectivity:
\begin{itemize}
    \item Helikites, a combination of kites carrying radio relays even at extreme conditions (of windspeed of 100 km/hr) without being severely affected by sea conditions. Helikites can be either tethered on land or sea platforms.
    \item Using the unused TV channels in the VHF and UHF bands for long-range line-of-sight (LoS) transmission. 
    \item  Multihop relaying for radio range extension of standard wireless communication such as universal mobile telecommunications system universal mobile telecommunications systems (UMTS) and LTE (long term evolution). 
\end{itemize}
The BLUECOM+ trials deployment reported single-hop and two-hop land-sea communications. The air-air links use IEEE 802.11g at 500 and 700 MHz, while air-sea links are based on UMTS or LTE. The two-hop testing involved two helikites tethered at 120 m altitude from two vessels \cite{BlueComPlusExperimentalUpdate}.\newline
LTE-maritime is another exciting project aiming to provide broadband connectivity at sea. Reports from the test-bed implementation of the LTE-maritime showed 10 Mbps communication at a shore-to-ship distance of up to 100 km using base stations located at high altitude regions on land \cite{jo2019lte}.  
\subsubsection{Satellites-Based Maritime Communication System}
\hfill\\
In addition to VDES, maritime communication systems primarily use satellites to provide a wider coverage than standard techniques that utilize microwave frequency bands in the 1-40 GHz band (see Fig.~\ref{fig:EMSpectrum} for different satellite frequency bands). Among various satellite constellations, some popular ones are Inmarsat, Iridium, and Thuraya. Inmarsat relies on a 14 geostationary earth orbit (GEO) satellite constellation operating in the L-band to provide near-global connectivity with relatively high data rates reaching up to 50 Mbps. Tapping on 66 L-band satellites on a low-earth orbit (LEO), the Iridium constellation provides voice and messaging global connectivity, including polar regions. With coverage in more than 160 countries, Thuraya provides voice and data coverage. Thuraya operates using 2 GEO satellites in the L-band. Given that a single GEO satellite can provide a coverage of more than $35\%$, Thuraya achieves beyond 70\% global coverage. The data rates of Thuraya using the ThurayaIP device are limited to 444 kbps.\newline
There are also many very small aperture terminal (VSAT)-based solutions that can offer nearly-global voice and internet coverage using LEO and GEO satellites  \cite{Itellian,MarLink,KHV}. Some VSAT service providers offer products with on-board data rates up to a few tens of Mbps.

A summary of terrestrial and space-based RF maritime communication systems and the various projects aiming to provide maritime broadband coverage is given in Table \ref{tab:MaritimeProjects}.\newline
\begin{table*}[t!]
\centering
\caption{\label{tab:MaritimeProjects} Summary of RF-based maritime communication systems and projects.}
\begin{tabular}{|m{55pt}|m{92pt}|m{90pt}|m{55pt}|m{150pt}|}
\hline
System&Technology and Band&Coverage&Max Data Rates&Use Cases\\
\hline
DSC&VHF, Maritime band&64 km&9.6 kbps&Maritime voice calling\\
\hline
AIS&VHF, Maritime band&64 km&9.6 kbps&Track and monitor vessels movements\\
\hline
NAVDAT&MF, 500 kHz\newline HF, [4, 6, 8, 12, 16] MHz &$\sim$ 500 km&18 kbps&Broadcasting of security and safety information from shore to ships\\
\hline
VDES&VHF&500 km&300 kbits& Establishing digital two-way communication between ships, satellite, and shore\\
\hline
TRITON&IEEE 802.16d, 5.8 GHz&shore-to-ship: $\sim$ 14 km \newline ship-to-ship: $\sim$8.5 km&6 Mbps& Providing offshore broadband internet access\\
\hline
Mare-Fi&IEEE 802.11n, 5.8 GHz&7 km&1 Mbps&Offshore WiFi connectivity\\
\hline
MariComm&LTE/WLAN 824-894 MHz\newline /(5.825, 5.785, 5.765) GHz&100 km& $>$1 Mbps&Providing broadband internet to ships \\
\hline
BLUECOM+ &Air-Air: IEEE 802.11g, 500/700 MHz\newline Air-Surface: LTE, 800  MHz &$>$100 km&$\sim$3 Mbps&Providing broadband internet access\\
\hline
LTE-maritime&LTE,\newline Uplink: 728-738 MHz\newline Downlink: 778-788 MHz&100 km&10 Mbps&Ship-to-shore communication\\
\hline
Inmarsat&GEO satellites,\newline Uplink: 1626.5-1660.5 MHz \newline Downlink: 1525-1559 MHz &Global, except polar regions&50 Mbps&Mobile and data services\\
\hline
Iridium&LEO satellites, Ku band&Global, except polar regions&46 Mbps&Providing on-ship voice calling and internet access\\
\hline
Thuraya&GEO satellites, L band&161 countries&444 kbps&Providing on-ship voice calling and internet access\\
\hline
VSAT&LEO and GEO satellites, Ku and Ka bands&Global, except polar regions&$\sim$10s Mbps&Providing on-ship voice calling and internet access\\
\hline
\end{tabular}
\end{table*}
\subsubsection{Aerial Networks for Maritime Communication}
\hfill\\
Aerial networks involve the use of UAVs, flying up to a few hundred meters above the sea surface and HAPs flying at the stratosphere, at least 20 km from the ground. Recent use of UAVs improves search and rescue missions by providing quick on-demand network deployment after disasters and also supporting mobility \cite{XianlingTCOM20}. In maritime networks, UAVs can relay the information sent from a ground station to mobile vessels beyond the LoS limit or when an LoS path is unavailable.
Moreover, UAVs can also help retrieve information from IoT devices located in the oceans and relay information to/ from unmanned surface vehicles (USVs). The main restriction of using UAVs is the limited flying time restricted by the carried load and the battery or fuel cell \cite{PANAppEng19}. Using a tethered UAV connected by an electrical cable connected to a power source can help relieve such limitations. Recent studies have shown the feasibility of using tethered UAVs fixed on buoys \cite{TetheredUAV21}. Tethered UAVs can hover in a certain place tens of meters above the seawater with limited coverage and mobility. In addition to the power cable, tethered UAVs can be connected with optical fibers to enable high transmission rates.
Due to their higher flying altitude compared to UAVs, HAPS flying at about 20 km from sea level offer an extended coverage radius that could reach hundreds of kilometers \cite{HAPSComst21}. The autonomy of HAPS can be as long as several months and, with fewer load weight restrictions, can carry large antennas \cite{BelmekkiIEEEOJVT22}. A recent demonstration by Airbus and NTT DOCOMO, INC. using their solar-powered Zephyr HAPS aiming to extend connectivity in the air and sea has reported connectivity in a range of up to 140 km \cite{Zephyr}. The HAPS-enabled connectivity to mobile terminals could allow users to use their mobile devices without needing a dedicated antenna.

\subsection{On-board Communications}
Another major aspect of maritime communications is the connectivity among different entities on the ship. Unlike conventional indoor radio wave propagation, radio wave propagation inside ships principally constructed with steel can be different. For instance, radios are used for communication between ship crew members, and a wireless sensor network may be used to monitor the movement of perishable and dangerous goods in shipping containers \cite{yingjun2010shipping}. Therefore, it is essential that the wireless channel for on-board applications is carefully modeled considering various scenarios.
Balboni \textit{et al.} reported a series of seminal investigations on radio channel characterization inside navy ships at a frequency range in the microwave band between 800 MHz and 2.6 GHz \cite{balboni2000empirical}.
The authors reported root-mean-square (RMS) delay spreads ranging between 70 and 90 ns \cite{balboni2000empirical} and path loss gradients ranging from 1/2 to unity. Path loss and the RMS delay spread were found independent of frequency over the considered frequency range. For on-board communications, channel measurements have been reported in different types of ships in various studies \cite{NoblesMILCOM03, MariscottiIMTCP10, MariscottiMeasurement11, MAOTCOM12}. The channel impulse responses inside compartments and within a passageway of a ship were obtained using a vertical network analyzer for 2 GHz and 5 GHz \cite{NoblesMILCOM03}. Channel sounding measurements were conducted in the restaurant hall and corridors of a cruise ship at 2.4 GHz \cite{MariscottiIMTCP10, MariscottiMeasurement11}. LoS and non line-of-sight (NLoS) channel measurements and 3D ray-tracing simulations were performed in the UHF band (from 225 to 450 MHz) inside a cargo hold of a merchant ship \cite{MAOTCOM12}, deriving the path LoS models for both scenarios. In \cite{mao2010study}, the channel characteristics and temporal fluctuations related to the propagation of VHF waves between the engine control room (ECR) and the bridge room have been examined in a vessel. Due to the dense multipath environment formed by the metallic structures inside the channel, broadband propagation may be impossible.
Furthermore, \cite{de2021radio} tested the channel in three separate places within the ship in a broader comprehensive investigation conducted in larger bands (868 MHz, 2.5, 5.25, and 60 GHz). The path loss exponents for sub-6 GHz were 1.21, 1.14, and 1.36 for 868 MHz, 2.4 GHz, and 5.25 GHz, respectively. However, the path loss exponent for mmWave wireless communications was more significant, i.e., 1.9.
\subsection{FSO and Hybrid RF/FSO for Maritime Communications}
\label{Subsec:FSO}
A particular advantage of FSO in maritime communication is the difficulty of interception and immunity to jamming contrary to RF signals, opening many military and civil applications opportunities. Various demonstrations were conducted mostly for military applications investigating the potential of FSO deployment in maritime military communication. Initial laser-based maritime communication demonstrations date back to 1970s \cite{GiannarisSPIE77}. A full-duplex heterodyne laser transmission was demonstrated using two 1600-nm CO$_{2}$ lasers over an 18.2 km maritime link in San Diego (California, US) within the optical convert communications using laser transceivers (OCCULT) experimental research initiative \cite{GiannarisSPIE77}. An automatic acquisition mechanism was involved with reciprocal pointing and tracking to ensure stable communication of the coherent two-way communication. 
In 2006, within the yearly Trident Warrior exercise, FSO systems were installed on two naval vessels \cite{RabinovichSPIE10}. A high-quality 300-Mbps uncompressed video transmission was reported over a maximum distance of 17.5 km in the Pacific, and the data link was transmitted with no disruptions or delay over 10 hours \cite{RabinovichSPIE10}. A bidirectional muti-Gbps FSO transmission was conducted off the mid-Atlantic coast between a tower on Cedar Island (Virginia, US) and a JHU/APL research vessel with varying distances between 2 and 22 km \cite{JuarezLSC10}. In 2017, a team from the Johns Hopkins University Applied Physics Laboratory (APL) demonstrated up to 7.5 Gbps FSO communication between two moving ships \cite{TW17Demonstration}. During the 14 hours up-time of the FSO terminal in a ship-to-shore configuration, data rates between 1 and 2 Gbps were reported for ranges exceeding 25 km. FSO link ensured voice communications for distances more than 35 km and sent messages up to the maximum available LoS distance of 45 km.

\begin{figure}[h]
    \centering
    \includegraphics[width=1\linewidth] {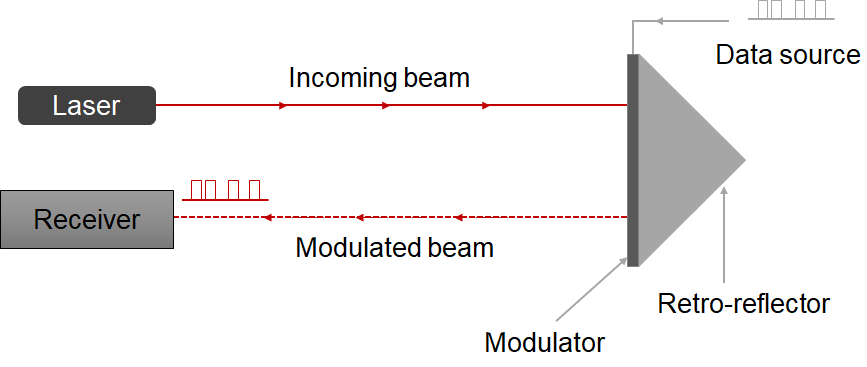}
    \caption {Diagram of an MRR-based FSO. The MRR couples a corner-cube retro-reflector and a modulator.  }
    \label{Fig:MRRModulation}
\end{figure}

Moreover, modulating retro-reflector (MRR)-based maritime links were demonstrated in \cite{MooreSPIE02,RabinovichSPIE05,BurrisSPIE09}. 
The diagram of an MRR is shown in Fig.~\ref{Fig:MRRModulation} that combines an optical retro-reflector with a modulator to reflect modulated optical signals (initially emitted by a laser interrogator) directly back to an optical receiver, allowing the MRR to function passively as an optical communication device without emitting its own optical power. MRRs are mainly used in maritime for FSO link characterizations.
FSO transmissions involving an array of MRRs with data rates ranging from 100 to 500 Mbps were conducted in the Chesapeake Bay (in the Mid-Atlantic region, US) over 32.4 km folded round-trip distance  \cite{MooreSPIE02}.
Using an array of 5 quantum-well-based MRR, a series of shore-to-boat FSO communications using a 1550-nm laser over a distance of 2 km in the Chesapeake Bay achieving data rates up to 5 Mbps \cite{RabinovichSPIE05}. 
A 32-km round-trip FSO communication using a 1550-nm laser modulated by an analog RF signal was reported through a maritime link using a retro-reflector at the Tilghman Island \cite{BurrisSPIE09}.\newline
\indent Beyond high-data-rate transmission and the use of MRRs, other experimental efforts involved deep investigations of the various propagation effects on FSO in maritime environments. For instance, the effects of turbulence and extinction on a 7.2 km FSO maritime path were investigated in \cite{GadwalSPIE06}. 
A summary of FSO trials and demonstrations is given in Table \ref{tab:FSODemonstrations}.
\begin{table*}[t!]
\centering
\caption{\label{tab:FSODemonstrations} Summary of FSO field trials and demonstrations.}
\begin{tabular}{|m{20pt}|m{20pt}|m{190pt}|m{190pt}|}
\hline
Ref.&Year&Demonstration description&Major outcomes\\
\hline
\cite{GiannarisSPIE77}&1977&- CO$_{2}$ lasers operating at 10.6 $\mu$m were used to build heterodyne systems for full-duplex ship-to-ship communication at close 5 MHz.\newline - Automatic acquisition and reciprocal pointing and tracking mechanisms were involved.&- Reciprocal pointing and tracking ensured stable communication.\newline - High-frequency stability is achieved. \\
\hline
\cite{RabinovichSPIE10}&2010&- A high-speed FSO transmission, in the form of a pre-taped video and live video feed between two US navy vessels, was established over 10 hours.
& - Up to 17.5 km transmission range and 300 Mbits were reported.\newline - No artifacts or delays in the videos were reported in clear weather transmission.\newline- During rain, slight video artifacts were obtained at ranges less than 10 km.\\
\hline
\cite{JuarezLSC10}&2010&- Two bidirectional ship-to-shore 1550-nm FSO field trials conducted off the mid-Atlantic coast near Wallops Island in July and September 2009\newline - The propagation distance was varied from 2 to 22 km, visual horizon.\newline - Two adaptive optics (AO) units were used to compensate for beam distortions and pointing errors. &- Up to 10 Gbit/s data rate was achieved.\newline - Daytime atmospheric turbulence is stronger than nighttime.\\
\hline
\cite{TW17Demonstration}&2017&- Demonstration of high-speed FSO transmission using terminals developed by APL Engineers in the 2017 Trident Warrior exercise.\newline - Demonstrations involved ship-to-ship and ship-to-shore communications. &- Up to a 7.5 Gbps transmission rate between two moving vessels was reported.\newline
- For 14 hours total up-time in ship-to-shore testing:\newline
* Error-free transmissions with data rates between 1 and 2 Gbps at more than 25 km ranges.\newline
* Voice communications for ranges up to 35 km.\newline
* Operational chat messaging at maximum LoS of 45 km.\newline
- Sea spray and fog were the major challenges.\\
\hline
\cite{MooreSPIE02}&2002&- MRR-based 32.4 km round-trip transmission using an array of 22 MRR.\newline - The height of the laser interrogator was 30 m from the water surface, while the height of the MRR from the surface was set to 15 to strengthen the propagation effects of the propagating laser beams. 
&- Data rates between 100 and 500 Mbps were demonstrated with a bit error rate (BER) below $10^{-5}$. \\
\hline
\cite{RabinovichSPIE05}&2005&- A series of FSO communication tests using a 1550-nm laser with data rates up to 5 Mbps over a 2 km distance from a ship to a boat in the Chesapeake Bay (Maryland, Virginia, US). \newline- Various weather conditions were covered over a one-year period&- Scintillation is the major challenge.\newline - At low and medium turbulence regimes, results are consistent with modeling proposed in \cite{AndrewsBook}.\newline - Experimental data are not in good agreement with theoretical models at a high turbulence regime.\\
\hline
\cite{BurrisSPIE09}&2009&- 32 km round-trip MRR-based FSO communication using a 1550-nm laser modulated by an analog RF signal through a maritime link.\newline
- Retro-reflector fixed at Tilghman Island.&- The analog modulation link was subject to turbulence, which could have been compensated if a digitizer had been used. \\
\hline
\cite{GadwalSPIE06}&2006&- Investigated the effect of turbulence and attenuation using a 1060-nm laser across a 7.6-km maritime path.& - Near-surface marine environment that is appropriate for ship-to-ship or ship-to-shore communications is an especially stressing propagation environment.\\
\hline
\end{tabular}
\end{table*}
FSO is not only restricted to horizontal links but can also be involved in vertical links, such as HAPS to vessel links. FSO is also a core technology to provide fiber-like data rates in satellite crosslinks in large LEO constellations aiming to provide wide maritime connectivity such as Telesat Lightspeed and SpaceX Starlink constellations \cite{Starlink,Telsat}.

Besides FSO-only links, various terrestrial hybrid RF/FSO schemes are also recently investigated in the literature \cite{TrichiliOJCS21, NadeemJSAC09, Gregorycharacterization11}. These hybrid links can be well suited for maritime communication between ships and ships to shore. Nevertheless, research on hybrid RF/FSO links for maritime networks is still in the early stages and needs further research.

\subsection{Challenges and Difficulties of Current Maritime Communication Technologies}
Having seen the different forms of maritime communication, here we summarize the various challenges for each technology. Starting with the RF-based solutions in the VHF band, namely, DSC and AIS, the available bandwidth and the limited range are the main limiting factors. A data rate of around 10 kbps over a few tens of km can only ensure half-duplex voice calling or sending and receiving safety and operation messages. The reach of extended WiFi technology is limited to a few kilometers off the costs (for example, the MareFi project \cite{MareFi14}). Cellular network access and broadband solutions based on WiMAX or LTE are restricted to near-shore locations, although in some cases can cover up to 100 km from the shore.
Aerial solutions such as the use of UAVs and HAPs are still limited in practice.  
The limited available bandwidth is also a constraint for satellite communication offered by any of the presented satellite systems presented in Table~\ref{tab:MaritimeProjects}. The high latency caused by the long propagation distance is another major challenge for maritime communication involving satellites in the GEO orbit. It should also be noted that satellite services are proprietary, making the cost of terminals and operation fees non-affordable for small vessels and fishing boats.\newline
FSO communication is constrained by turbulence and weather conditions that could be severe, affecting the link availability. The Earth curvature LoS limit also defines the maximum distance FSO links can achieve in an ideal medium. Maintaining stability between the terminals is challenging.
A summary of the challenges of maritime communication is given in Table~\ref{Tab:TechnologiesChallenges}. We note that we did not discuss the challenges of aerial networks as their use in maritime communication is still limited in practice, and their future potential and challenges will be covered in a later section of the manuscript.

\begin{table*}
\caption{\label{Tab:TechnologiesChallenges}Summary of Challenges and Difficulties of Maritime Communication Technologies.}
\centering
\begin{tabular}{|p{3cm}|p{9cm}|}
\hline
Technology &Challenges \\
\hline
RF&- Limited bandwidth \newline
- HF, MF, and VHF solutions are restricted to half-duplex voice communication\newline
- Limited coverage \\
\hline
Satellite&- High latency and low data rate\newline
- High cost of terminal and monthly/yearly service fees \newline
- Subject to Multipath\\
\hline
FSO&- Sensitivity to maritime turbulence and weather condition\newline 
- Maintaining alignment between the communicating terminals\newline
- Maximum theoretical distance defined by the Earth curvature\\
\hline
\end{tabular}
\end{table*}
\section{Building Blocks of Maritime Communications}
\label{Sec:BBlocks}
In this section, we discuss the fundamental physical layer aspects of maritime communications, including channel modeling, modulation and coding for RF, FSO, and hybrid systems. We also overview other key performance parameters, such as coverage and capacity and radio resource management.
\subsection{Channel Models (RF/FSO/Hybrid Systems)}
Various reports in the literature have studied maritime channel effects and modeling for RF, FSO, and hybrid technologies. In the following, we present the different channel models studied in the literature.
\subsubsection{RF-based Channel Models}
\begin{figure}[h]
    \centering
    \includegraphics[width=1\linewidth] {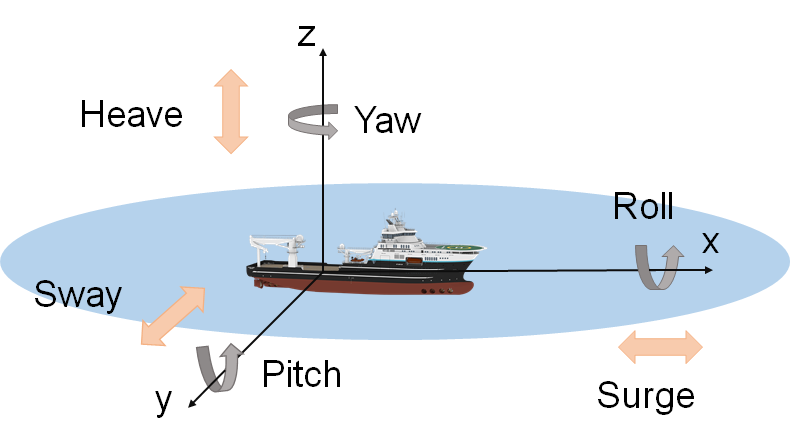}
    \caption {Vessel linear and rotational motions in the sea.}
    \label{seamotion}
\end{figure}

\begin{figure}[h]
    \centering
    \includegraphics[width=\linewidth] {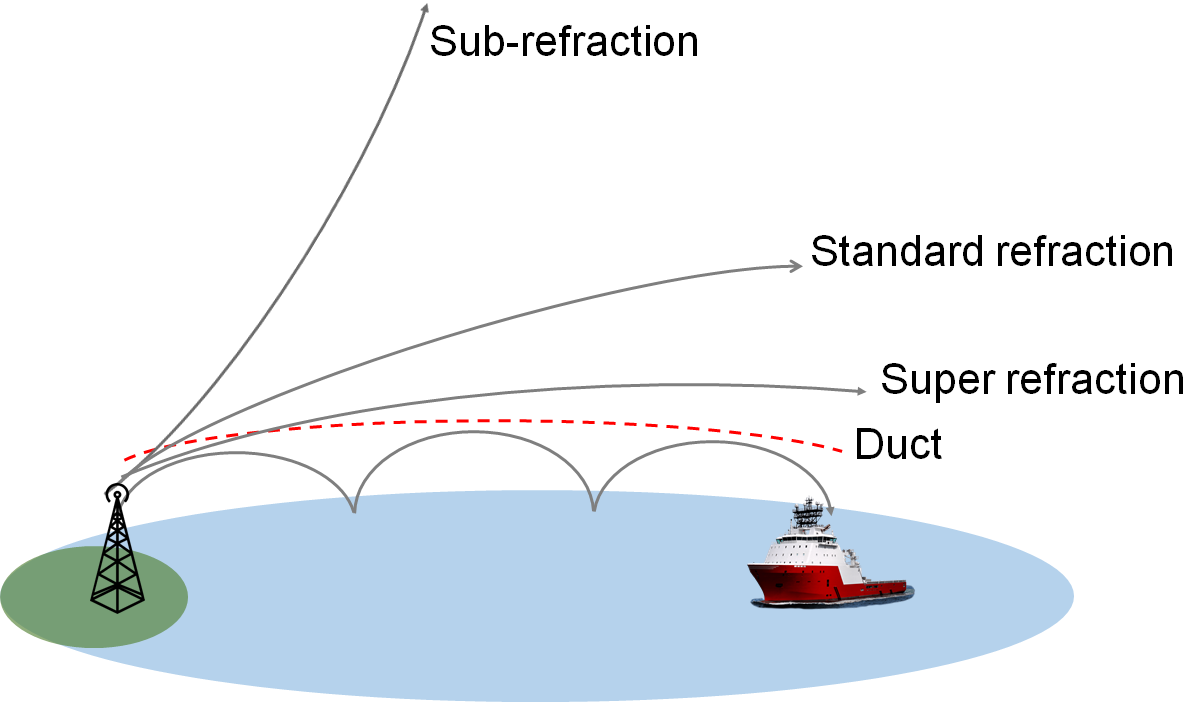}
    \caption {Ray propagation under different atmospheric refraction conditions. The duct is a horizontal layer that tends to follow the Earth's curvature. } 
    \label{ducting}
\end{figure}

\hfill\\ 
The characteristics of maritime communication channels are different from conventional terrestrial wireless channels. The difference is mainly due to the following features: sparsity, wave-induced instability, and ducting phenomena.
\newline
\textbf{Sparsity}: Since RF-based maritime channels do not suffer from scattering, the assumption of Rayleigh fading is no longer valid. Therefore, finite-scattering models introduced for terrestrial RF communication \cite{bajwa2009sparse, saleh1987statistical} can be used for maritime channel modeling. Sparsity in maritime communication also manifests in user distribution since users are broadly distributed in the sea.   \newline
\textbf{Wave-Induced Instability:} The movement of waves causes periodic changes in the height and orientation of the on-board antennas that lead to a reduction in the received message power. The movement of the waves can be considered as linear motion, or rotation motion \cite{zhao2014radar, reddy2016analysis}. The linear motion exists along one specific axis (x-only, y-only, or z-only). In contrast, the rotation motion considers the movement along all three axes (see Fig.~\ref{seamotion} for possible vessel motions in the sea). Aside from the link mismatch effect due to the sea wave movement, water motion also leads to radio transmission scattering, particularly at the air/water interface. Three metrics are often employed to characterize sea wave movement: crest-to-trough wave height (difference between the highest surface part of a wave and the lowest part), wave wavelength, and wave period \cite{hubert2012impact, huang2015maritime}.\newline
\textbf{Evaporation Ducting Phenomena:} The refractive index of the atmosphere changes with height, and the refractivity variations in the lower layer of the atmosphere depend on wind, temperature, pressure, and, most importantly, humidity, leading to duct formation  \cite{ErginCommag14,ErginTAP15}.
For instance, as shown in Fig.~\ref{ducting}, there are four possible refractive conditions: the sub-refraction, standard refraction, super-refraction, and evaporation refraction. In the evaporation refraction condition, the signal is \textit{trapped} inside the ducting layer and refracted back by the duct to the sea surface \cite{ErginCommag14}. The evaporation ducting phenomena is almost permanent in coastal and maritime locations, and the duct height ranges between 10 and 20 m, with a maximum of 40 m \cite{ErginCommag14}.  \newline
The sparsity, wave-induced instability, and ducting can occur for different wireless links, i.e., ship-to-ship, air-to-ship, shore-to-ship, and satellite-to-ship. 
In the following, we examine these RF-based wireless links.\newline
\begin{figure}[h]
    \centering
    \includegraphics[width=1\linewidth] {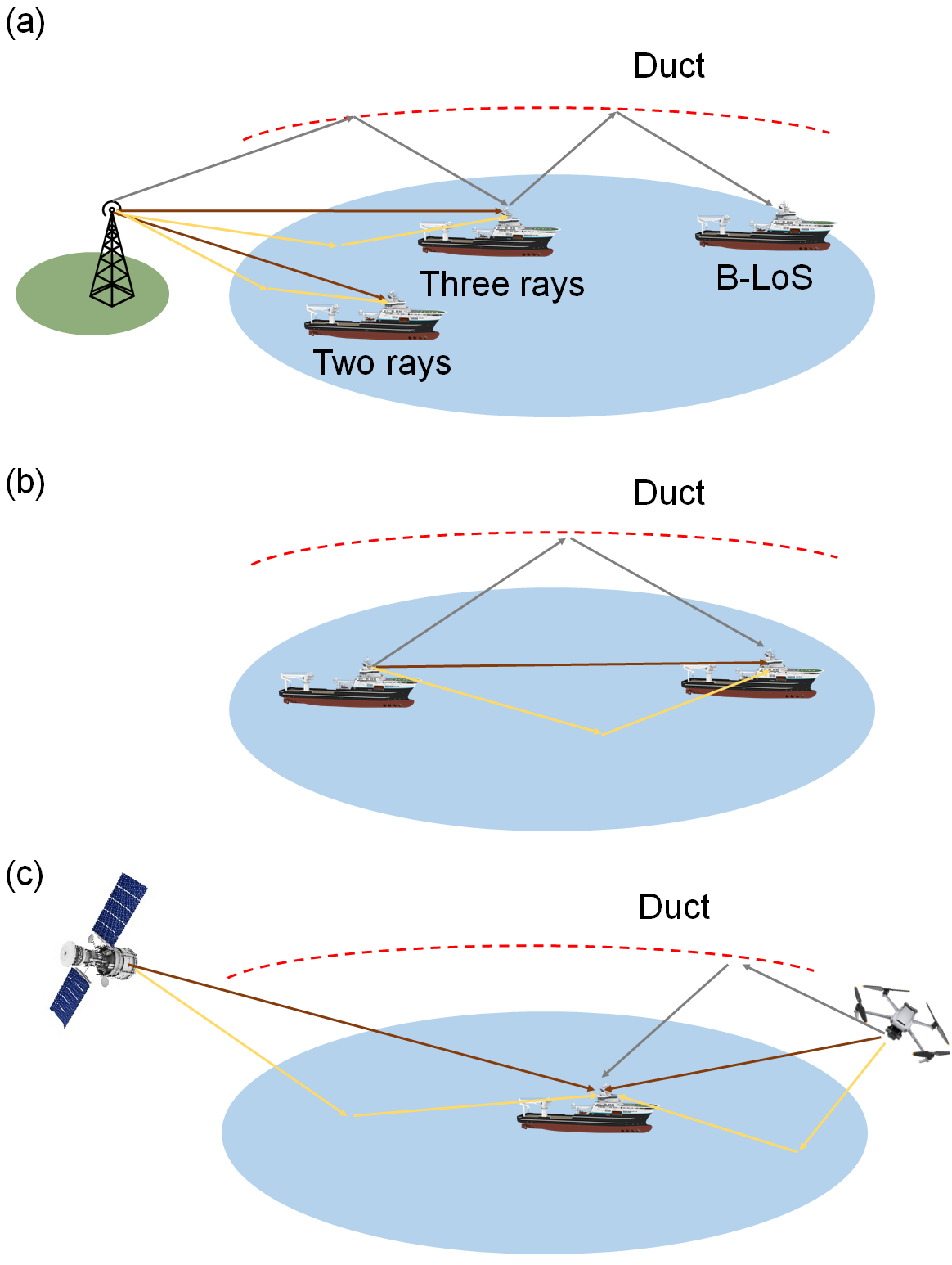}
    \caption {Illustration of signal propagation in (a) shore-to-ship, (b) ship-to-ship, and (c) satellite/air-to-ship communication scenario.}
    \label{Shore/Satellite-to-ship}
\end{figure}
\indent \textbf{Shore/Ship-to-Ship Links}:  shore-to-ship and ship-to-ship links are primarily distance-dependent, as shown in Fig.~\ref{Shore/Satellite-to-ship}. In the case of short-range, the wireless channel acts as a two-ray model by taking into account the earth's curvature, and the channel response can be formulated as follows:
\begin{equation}
h_{2}(t, \tau)=\delta(\tau- \left.\tau_{0}(t)\right) +\alpha_{s}(t) \exp \left(j \varphi_{s}(t)\right) \delta\left(\tau-\tau_{s}(t)\right),
\end{equation}
where $\alpha_{s}(t)$ is the amplitude of the surface reflection wave, $\varphi_{s}(t)$ is the phase difference between the direct route and the reflection wave, and $\tau$ is the propagation delay. $\alpha_{s}(t)$ may be modified by characteristics such as the reflection coefficient, shadowing factor, divergence factor, and surface roughness factor, whereas $\varphi_{s}(t)$ can be estimated geometrically using the curved earth approximation \cite{matolak2016air}. Mostly, for short distances between the transmitter and receiver, if the antenna is mounted at a good height, the maritime channel has LoS and NLoS reflected wave components, leading to a two-ray channel model assumption \cite{garroppo2008wimax}. On the other hand, the local scattering around the user in the ship cannot be neglected for the low-height antennas. Then we need to take more paths into consideration, which leads us to the so-called two-waves with diffusion power model introduced in (\cite{durgin2002new}, Eq.~(4)):
\begin{equation}
\tilde{V}=\sum_{i=1}^{N} V_{i} \exp \left(j \Phi_{i}\right)+X+j Y,
\end{equation}
where $N=2$, $V$ is the total voltage induced at the receiver antenna, which is composed of two components: the specular component $\sum_{i=1}^{N} V_{i} \exp \left(j \Phi_{i}\right)$ and the diffusion component $X+jY$, which has a complex Gaussian distribution and represents the sum of several individual weak waves. This model has been found to fit outdoor mmWave channels at 28 GHz \cite{romero2016fluctuating, romero2017fluctuating}.\newline
In the case of medium-range communication, a three-ray model can be used to characterize the channel as follows:
\begin{equation}\label{threeray}
    h_{3}(t,\tau)=h_{2}(t, \tau) +z_{3}(t) a_{3}(t) \exp \left(j \varphi_{3}(t)\right) \delta\left(\tau-\tau_{3}(t)\right), 
\end{equation}
where $a_{3}(t)$, $\tau_{3}(t)$, and $\varphi_{3}(t)$ denote the third multipath component's time-varying amplitude, propagation delay, and phase shift, respectively \cite{matolak2016air}. In Eq.~\eqref{threeray}, $z_{3}(t)$ is created by a random process that determines the chance of the third multipath component. \newline
In the long-range propagation case, a duct-based link could be the only way to establish the communication, as can be seen in \ref{Shore/Satellite-to-ship}(a). The duct act as a dielectric waveguide that can guide waves beyond LoS (B-LoS). B-LoS communication is very common in maritime, and many efforts have been conducted to understand and study its behavior. First, there were studies on wave refractivity, and propagation in the ducting layer \cite{ErginCommag14}. Authors of \cite{ErginTAP15} then developed a statistical large-scale path-loss model that helps to evaluate the path loss exponent and the propagation range using parabolic equation (PE) simulation, which is the approximation of Helmholtz wave equation \cite{sirkova2012brief}. PE is often solved numerically using one of three methods: split-step Fourier (SSF), finite difference (FD), or finite element method (FEM) \cite{sirkova2012brief}. The most appropriate strategy to solve the PE for a given situation is highly dependent on the scenario and set of circumstances. The SSF approach relies on fast Fourier transforms. As a result, PE with SSF is more computationally efficient than other approaches, and the methodology may provide precise and stable answers. The FD method gives the maximum resolution in simulating the boundary conditions by applying the Crank Nicholson finite difference technique. The FEM enables more accurate modeling of quick shifts in atmospheric conditions and more modeling flexibility for complicated boundary conditions. A comprehensive examination of these strategies can be found in \cite{sirkova2012brief}. There are also wave propagation tools that use these numerical models to solve the PE \cite{ErginTAP15}.\newline
\textbf{Satellite/Air-to-Ship Links}: 
The sea surface mainly causes multipath, creating two possible paths; an LoS path and another one reflected from the duct, sea surface, or both, as in the case of Air-to-Ship links, illustrated in Fig. \ref{Shore/Satellite-to-ship}(c). In terms of channel model, various experiments demonstrated that the Rician model is the most suitable statistical model for the wireless channel in satellite/air-to-ship communication links \cite{XianlingTCOM20,hagenauer1987maritime,wang2019doppler}. Nevertheless, recent works have tried to improve the communication performance between the UAVs and the ship, which requires more complex channel modeling with 2D or 3D formulation \cite{shi2014modeling,liu2021novel}. For example, in \cite{shi2014modeling}, 2D and 3D sea surface simulations are used to understand the signal propagation for UAV-to-ship links. For 2D, the finite difference time domain (FDTD) was used with a maximum 10 m variation. An alternating-direction implicit finite-difference time-domain (ADI-FDTD) was used for the 3D modeling. Although the 3D formulation needs high computational power, it is more realistic than a 2D simulation. Similarly, the authors of \cite{liu2021novel} considered 3D channel modeling by considering the multi-mobility of UAVs and the ship's motion at arbitrary speeds and directions. This approach helps to study some of the channel statistical properties in maritime communication between UAVs and ships, considering the mobility, speed, and clusters between the transmitter and the receiver. 
Due to the relative motion of satellites, UAVs' mobility, and ships' movement, the Doppler shift is a common issue for these links. Consider a typical LEO satellite at an elevation of 650 kilometers above the earth's surface, the Doppler shift may range between -4 and 4 kHz depending on the relationship between relative velocity, angle, and carrier frequency \cite{wang2019doppler}. Hence, various approaches have been proposed to estimate the Doppler shift with a particular focus on space-based AIS signals \cite{ChoiICTC13,MaICOACS16,MengOL18,MengWPC18,WangIEEETVT21}.
\subsubsection{FSO Channel Modeling}
\hfill\\
Like terrestrial propagation, optical beams propagating through maritime environments are subject to various weather conditions and turbulence. Weather conditions (including haze, humidity, fog, rain, etc.) cause attenuation, with effects lasting from a few minutes to several hours. Turbulence, however, causes effects with a timescale of a few milliseconds known as scintillation and beam wandering. Scintillation is the rapid random intensity fluctuations quantified by the so-called scintillation index, which is the variance of the irradiance fluctuation normalized by the square of the mean irradiance. Beam wandering is the random movements of the incoming laser beam on the receiver plane.\newline
\indent FSO power attenuation as a function of the propagation distance $z$ can be described by Beer's law as follows:
\begin{equation}
P(z)=P_{0}\exp(-\gamma(\lambda)z),
\end{equation}
with $P_{0}$ is the initial power, $\gamma(\lambda)$ is a wavelength-dependent attenuation coefficient, and $\lambda$ is the operation wavelength. $\gamma(\lambda)=\alpha+\beta$ is the contribution of two phenomena; absorption ($\alpha$) and scattering ($\beta$). Gaseous molecules and aerosol particles cause the absorption of the light phenomenon in the atmosphere. Coefficient $\alpha$ can be neglected for maritime FSO since FSO wavelengths are in the non-absorption atmospheric windows. There are three scattering types; Rayleigh, Mie, and non-selective scattering, and $\beta$ can be the contribution of all these three forms. Rayleigh is an all-direction scattering caused by particles smaller than the optical wavelength. The effect of Rayleigh scattering is negligible for wavelengths beyond 800 nm; therefore, its impact can be neglected for FSO maritime links incorporating IR laser sources. Mie scattering is mainly caused by particles with sizes comparable to the optical wavelength and originates from the fog. There are various empirical models in the literature for Mie scattering. Kim and Kruse's models are the most accepted ones. The scattering coefficient, $\beta_{Mie}$ [km$^{-1}$] in both models can be written in the following form:
\begin{equation}
\beta_{Mie}(\lambda)=\frac{3.91}{V}\left(\frac{\lambda}{\lambda_{0}}\right)^{-q},
\end{equation}
where $V$ is the visibility, $\lambda_{0}$ is a reference wavelength (commonly fixed at 550 nm), and $q$ is the size distribution of the scattering particles. For the Kruse, parameter $q$ is given as follows:
\begin{equation}
q_{Kruse} =
\begin{cases}
& 1.6~~~~V>50~\text{km} \\ 
&1.3~~~~6~\text{km}\leq V<50~\text{km}.\\
&0.58V^{1/3}~~~~V<6~\text{km} 
\end{cases}
\end{equation}
At low visibility Kim model provides higher accuracy:
\begin{equation}
q_{Kim} =
\begin{cases}
& 1.6~~~~V>50~\text{km} \\ 
&1.3~~~~6~\text{km}\leq V<50~\text{km}\\
&0.16V+0.34~~~~1~\text{km}<V\leq6~\text{km}. \\ 
&V-0.5~~~~0.5~\text{km}<V\leq1~\text{km}\\
&0~~~~V\leq0.5~\text{km}
\end{cases} 
\end{equation}

Non-selective scattering is caused by particles larger than the optical wavelengths, including rain and snow. Empirical models for rain and snow determined for terrestrial FSO links are applicable in maritime environments. For the rain model, the scattering coefficient can be be expressed as $\beta_{rain}=KR^{\alpha}$ [dB/km] with $R$ being the precipitation intensity in [mm/hr] and $(K,\alpha)$ are model parameters \cite{WirelessComBook}. For the snow, the scattering coefficient can be expressed as $\beta_{snow}=aS^{b_{s}}$ [dB/km] with $S$ being the snowfall rate [mm/hr] and ($a_{s}$,$b_{s}$) are snow parameters that take different values in wet and dry snow \cite{WirelessComBook}.  \newline
\indent There are various numerical and stochastic models to model the effect of turbulence in terrestrial FSO channels \cite{TrichiliJOSAB20}. Numerical models derived from the Kolmogorov turbulence theory can be used to model the random variations of the refractive index of the atmosphere that cause turbulence \cite{AndrewsBook}. A key parameter in the different numerical models is the refractive index structure parameter, $C_{n}^{2}$, which is a measure of the strength of fluctuations and could take values ranging from 10$^{-13}$ for strong turbulence and 10$^{-17}$ for weak turbulence.\newline
\indent Stochastic models for terrestrial atmospheric turbulence involve the lognormal distribution for weak turbulence, negative exponential distribution for strong turbulence, the Gamma-Gamma distribution for weak to medium turbulence, and the generalized Malag{\`a} model covering a wide range of turbulence strengths, \cite{Comst14}. The rich literature on terrestrial FSO channel modeling \cite{AndrewsBook,Comst14} cannot be used to describe the propagation in a maritime channel mainly because turbulence shows different behaviors between terrestrial and marine environments \cite{GrayshanMarineModelling,ToselliMarineModelling}. \newline 
An early study by Friehe \textit{et al.} revealed that $C_{n}^2$ fluctuations over the water are different from the inland case, mainly due to the significant humidity variations \cite{FrieheJOSA75}.
However, there have been equally several experimental and theoretical efforts on marine FSO channel modeling. The authors of \cite{GrayshanMarineModelling} introduced a novel marine turbulence spectrum and derived a theoretical expression for the irradiance fluctuation under a weak turbulence regime.\newline
An experimental study conducted at the Piraeus Port (Greece) proposed a novel empirical model for attenuation dependent on three parameters; the relative humidity, the atmospheric temperature, and the wind speed \cite{PiraeusModel}. The experimental validation was based on a $\sim$3 km FSO link operating at the 850 nm wavelength with transceivers fixed 35 m from sea level. The experiment reported in \cite{JuarezLSC10} revealed that daytime atmospheric turbulence is stronger than nighttime turbulence.\newline
Li \textit{et al.} investigated the BER performance of a coherent FSO employing a quadrature phased shift keying (QPSK) modulation subject to maritime atmospheric turbulence and considered compensating the effect of turbulence distortions using an AO unit \cite{CvijeticAO15}. The authors showed that using AO to compensate for beam distortions could significantly improve the BER system performance by several orders of magnitude \cite{CvijeticAO15}.\newline
\indent Further studies investigated the propagation of light beams with complex light structures through oceanic environment \cite{ZhuJOSAA16}. Note that spatially structured light beams are used to increase the transmission capacity by multiplexing multiple orthogonal light modes in the same beams (\cite{TrichiliComst19}. Authors of \cite{ZhuJOSAA16} studied the propagation dynamics of partially coherent modified Bessel-Gaussian (that carry OAM) beams in an anisotropic non-Kolmogorov maritime atmosphere and derived an analytical formula on the evolution of the powers of the received beams.\newline

\subsubsection{Hybrid Models}
\hfill\\
Experimental investigations revealed that RF and FSO are affected differently by weather conditions \cite{NadeemJSAC09,KimSpie11}. For example, FSO links are highly sensitive to fog and snow but resilient to rain. In contrast, RF links are resilient to fog and snow but severely degraded by rain. Nadeem \textit{et al.} studied the impact of dense maritime fog (with measurements collected in La Turbie (Nice, France)) on an FSO link operating at 850 nm installed as the primary link with an RF backup link operating at a frequency of 40 GHz \cite{NadeemJSAC09}. The authors reported a significant deterioration of the FSO link by 480 dB/km attenuation. However, a $100\%$ availability of the hybrid system was reached thanks to the low RF signals attenuation at 40 GHz. Gregory and Badri-Hoehe conducted a 6-month measurement campaign on a 14-km long RF/FSO hybrid link with the FSO system operating at a 1550-nm wavelength and the RF system operating at 38 GHz \cite{Gregorycharacterization11}. The main motivation of the work was to correlate the hybrid link results with the weather conditions that were measured simultaneously. The RF link exhibited more than $99\%$ availability over the measurement period and was mainly degraded by rain. The FSO link was affected primarily by fog leading to severe attenuation, especially in the daytime. The authors equally provided the max and average values of the Fried parameter. We note that the Frier parameter is a measure of coherence length, collected over the measurement period to study the effect of scintillation, and is also helpful in applying turbulence mitigation strategies such as using multiple-input multiple-output (MIMO) FSO. In a MIMO FSO configuration, transceivers separated by $r_{0}$ can ensure diversity, and each path can experience different turbulence effects.

\subsection{Modulation and Coding Schemes}
Besides channel modeling, maritime network modulation and coding schemes are other essential physical layer issues. We will first start by presenting the modulation and coding techniques used in maritime communication for RF and then for FSO-based systems.

\subsubsection{RF-based schemes}
\hfill\\
Multiple modulation and coding techniques have been proposed for theoretical and experimental reports in the literature \cite{lazaro2019vhf,gamache1999oceanographic,hagenauer1982data}. For instance, the authors of \cite{lazaro2019vhf} proposed an adaptive coding and modulation (ACM) for the VHF band-based VDES (See Section \ref{Subsec:RFTech}). The ACM consists of dynamically changing the modulation format and the coding rate according to the experienced signal-to-noise ratio (SNR). When a ship is far from shore, the received signal may be weak and slightly higher than the background thermal noise, leading to a low SNR \cite{lazaro2019vhf}. In such a case, the communication should involve a robust modulation such as the $\pi/4$QPSK and a channel code rate of $1/2$. When the ship is close to shore, leading to higher SNR, 16-quadrature amplitude modulation (QAM) with a rate $3/4$ channel code may be used \cite{lazaro2019vhf}. The author of \cite{gamache1999oceanographic} uses an oceanographic data link (ODL) system, which is a two-way connection: a forward link (from the hub to the terminal) and a return link (terminal to hub). This bidirectional feature enables dynamic experimentation and remote sensor system monitoring and control. The ODL architecture offers several methods for multi-access, including direct sequence spreading to avoid interference from neighboring satellites, TDMA, FDMA, and CDMA. The ODL system's access protocols may be customized to a particular network and can handle thousands of users per MHz for oceanographic applications. In an earlier paper \cite{hagenauer1982data}, the authors utilized a stored channel method, simulating three different links, i.e., satellite-to-ship, buoy-to-satellite, and base station-to-land mobile. The satellite-to-ship link uses a binary phase shift keying modulation scheme, whereas the base station-to-land mobile link utilizes a variety of modulation schemes like PSK, differential phase shift keying (DPSK)/FM, PSK/FM, and digital FM.\newline
\subsubsection{Optical Related Schemes}
\hfill\\ 
There are two types of FSO systems: intensity modulation/direct detection (IM/DD) and coherent systems. IM/DD systems are relatively simple and consist of modulating the intensity of a laser and directly detecting light signals at the reception by a photodetector. IM/DD can only support unipolar modulation, including on-off shift keying (OOK), pulse amplitude modulation (PAM), and pulse position modulation (PPM). Coherent systems provide phase tracking by a so-called local oscillator (LO), enabling encoding information using complex multilevel modulation formats like QPSK and M-array QAM. The incoming information-carrying signal is mixed with the LO at the receiver. Compared to IM/DD, coherent systems generally allow for better background and shot noise resilience and ensure the transfer of higher data rates, but at the expense of cost and complexity.\newline
\indent There are two types of coherent detection depending on the frequency of the LO; homodyne and heterodyne. In homodyne detection, the LO frequency matches the laser frequency. Mixing the LO and the information signal in heterodyne detection yields a signal in the microwave region. The demonstration reported in 1977 is an example of a coherent heterodyne maritime optical communication with FM modulation \cite{GiannarisSPIE77}. In 2005, a 5.62 Gbps homodyne FSO transmission was conducted between two Canary Islands (two ground stations on La Palma and Tenerife Islands) separated by 142 km mostly above the sea and incorporating a BPSK modulation \cite{CanaryTransmission}. The use of BPSK modulation exhibited robustness to atmospheric conditions \cite{CanaryTransmission}. Authors of \cite{CvijeticAO15} evaluated the effect of maritime turbulence on a QPSK coherent system. They found that the maritime FSO system has a higher BER when compared to a terrestrial one experiencing the same turbulence strength.\newline
\indent Authors of \cite{qiao2021performance} proposed using DPSK modulation with repetition coding consisting of sending the same message several times to benefit from time diversity. When compared to other modulation formats (PPM, PAM, OOK, and QPSK) through BER simulation analyses, the authors found that DPSk can provide a good compromise between long-distance and system capacity. DPSK can equally solve the phase ambiguity condition in BPSK modulation \cite{qiao2021performance}. It was also found that by increasing the repeat time, repeating coding can significantly suppress BER without the need for aperture averaging used commonly at the receiver to undo the effect of atmospheric turbulence \cite{qiao2021performance}. 
Another way of modulation in FSO links is through the use of MRRs as reported in \cite{MooreSPIE02,RabinovichSPIE05,BurrisSPIE09} and discussed in Section \ref{Subsec:FSO}.\newline
\indent In the case of an MRR-based FSO, the modulation of the beam emitted by the laser is conducted at the MRR through a modulator. The modulated light beam is reflected back to a receiver through a retro-reflector \cite{RabinovichSPIE05} (See Fig.~\ref{Fig:MRRModulation}). We stress that the MRR is a passive component and does not emit light. The use of MRR is not restricted to maritime FSO experiments. Still, it has been widely used in terrestrial and Earth-satellite lasercom links for link characterization and ranging purposes (\cite{RabinovichSPIE05}, and references therein). The modulation at the MRR can be ensured using different technologies such as liquid crystals \cite{FLC} and quantum well \cite{RabinovichSPIE05}.\newline
\subsection{Coverage and Capacity}
Coverage and capacity are two other performance metrics that many studies tried to increase for maritime communication networks. VDES and NAVDAT provide the highest maritime coverage. However, MF, HF, and VHF bands are limited in terms of the achievable data rates restricted to a few tens of kbps. Going to higher frequencies can improve the capacity of the RF maritime links, as shown by reports from the deployment of the Korean LTE-maritime project that demonstrated a coverage of 100 kilometers from shore with a data rate in the order of a few Mbps \cite{jo2018validation}. Using the evaporation duct to extend the reach of microwave signals beyond the LoS can extend maritime communication coverage, as shown by a seminal study in \cite{WoodsIEEEJO19} reporting a 10 Mbps data rate over a 78 km link between the Australian mainland and the Great Barrier Reef (GBR) (off the east coast of Queensland, Australia) at a frequency of 10.6 GHz. Three long-range microwave duct-based links with distances of 109, 63, and 87 km have been recently demonstrated in the South China Sea \cite{MaIEEECommag22}. \newline
MIMO can be an effective solution to boost the capacity of maritime RF. A MIMO antenna design has already been tested for maritime communications in the 5.1 and 5.9 GHz bands, as recently reported in \cite{MaritimeMIMO}. Multihop relying can also extend maritime coverage, as demonstrated by the test conducted within the MariComm project \cite{MariComm15}.\newline
Space-based maritime communication can have considerably broader coverage compared to VHF, and other RF-based solutions \cite{BekkadalITST09}. However, satellite-based MCNs can have blind zones and covered areas subject to severe interference \cite{WeiAccess20}. To cope with these challenges, an environment-aware system was proposed in \cite{WeiAccess20} to optimize satellite-ground integrated maritime communication capacity.  
In the case of FSO, there are several approaches to increasing the coverage and capacity of the optical band. The FSO LoS is restricted by the Earth's curvature but may reach more than 30 km distances with terminals installed on typical ships. For instance, as reported in Table \ref{tab:FSODemonstrations}, voice communications can range up to 35 km, while operational chat messaging can reach up to a maximum LoS of 45 km \cite{TW17Demonstration}. Also, in terms of capacity, a maximum of 10 Gbps data rate was reported in \cite{JuarezLSC10}, and 7.5 Gbps was achieved between two moving vessels \cite{TW17Demonstration}. Extending the coverage of FSO beyond the LoS requires relays, and the amplify and forward scheme can be the most straightforward approach to accomplishing it. The current data rates provided by maritime FSO at tens of kilometers distance are enough to perform maritime communication operations and improve the connectivity to those sailing on board. But for applications requiring more capacity, using multiple wavelengths can significantly increase the transmission rates.
\subsection{Radio Resource Management}
Like in terrestrial wireless communication systems, RRM is essential for MCNs to manage the radio resources and other transmission characteristics at the system level. In \cite{mroueh2015radio}, the authors proposed a maritime Mobile Ad-hoc Network (MANET) with LTE nodes where the nodes represent ships forming a naval fleet headed by a shipmaster. This naval fleet is treated as a cluster in MANET, where the ship crew is referred to as a cluster node (CN), and the shipmaster is referred to as a cluster head (CH) \cite{yu2005survey}. Then, the allocated bandwidth is optimized at the CH to service all CNs while reducing the chance of the CH running out of radio resources \cite{mroueh2015radio}. In summary, this study aims to calculate the necessary bandwidth to be provided at the CH to accommodate all active CNs' traffic. They considered two transmission schemes: (a) the single input single output (SISO) configuration; and (b) the $2\times2$ MIMO setup with properly spaced antennas. They also demonstrated that compared to the SISO scenario, the $2\times2$ MIMO spatial multiplexing modes with a full diversity transmission mode for long-distance marine communication enhance the average network spectral efficiency and resource outage probability. Because of the complexity and high computation of the calculation in \cite{mroueh2015radio}, the authors also investigated the same problem, but this time by taking the ITU marine path-loss \cite{itu2013method} with 10\% time as a reference model \cite{kessab2016impact}. In \cite{duan2019joint}, the authors proposed to use MIMO antennas with a coastal two-hop relaying system to manage and communicate with users and boats in the offshore region. Although the boats in the offshore area are few, they may be grouped into many clusters consisting of multiple vessels close together. Their numerical results show that the algorithm is power-efficient, which may fit with the challenges faced in maritime communication. Additionally, it is worth noting that some recent research works use USV in their calculations to increase power efficiency and coverage \cite{zeng2021joint, zeng2021joint1}.

\subsection{Energy Efficiency}
Aside from boosting the maximum throughput and reach that could be attained, there has been some focus on energy saving to provide sustainable solutions for maritime networks. For instance, a software-defined networking (SDN) solution may reduce network latency and energy consumption while ensuring network flexibility and stability. In this context, the delay-tolerant networking (DTN) as a scheduling mechanism can be utilized in maritime communication with SDN as controller. Both offer a cost-effective solution, as discussed in \cite{yang2021efficient}. The authors of \cite{yang2021efficient} modeled DTN and SDN as a trade-off optimization problem and showed that modifying various factors could achieve a delay-energy trade-off. 
An energy-aware solution for marine data acquisition via IEEE 802.11-based wireless buoys network, known as Wi-Buoy, is proposed in \cite{ReddyAccess21}. The authors proposed an optimization framework to reduce energy consumption at a backhaul buoy source used to transfer seismic data collected from sensors on the seabed to a seismic survey vessel \cite{ReddyAccess21}. Motivated by the fact that most vessels follow designated shipping lanes, authors of \cite{WeiIEEETVT19} proposed using the shipping lane information to minimize the average power consumption. \newline
Recently, authors of \cite{HassanArxiv22} addressed the energy efficiency aspect of a maritime network from the perspective of UAVs used as a part of a space-air-sea non-terrestrial network connecting user equipment to LEO satellite using UAV-based aerial base stations.\newline
Schemes of self-powered maritime networks have also been proposed in the literature \cite{FahrajiUSNC19,FahrajiINFOCOM20}. Authors of \cite{FahrajiINFOCOM20} presented a self-powered marine communication system composed of many buoys. Each has a communication unit and energy-harvesting unit that harvests energy from the floating buoy movements due to ocean waves. The communication unit consists of a wireless router operating on the TV white space band. Deployment and simulation results showed that the proposed design could produce more energy than needed for the system operation \cite{FahrajiINFOCOM20}.
\subsection{essons Learned}
In each technology used for maritime communications, the signals are subject to channel conditions that differ from the terrestrial links. Many efforts have been made to model the RF, FSO, and space-based maritime communication channels. Various statistical channel models have been proposed for RF communication which mainly depends on the ship-to-ship or shore-to-ship distance. FSO channel modes depend on oceanic turbulence and weather conditions. Although many studies are available in the literature on maritime channel modeling compared to terrestrial networks, further efforts on modeling based on statistical analysis and channel sounding are needed. The coverage and capacity of different maritime communication technology are improving. Still, given the vast oceans and the growth of the marine industry, the current technologies are limited in coverage and capacity. Involving satellites can extend the coverage and, in some cases, the capacity; however, the cost is still an issue and may not decrease due to the high cost of launching satellites into orbits. Energy efficiency in maritime networks has been addressed in different aspects spanning from using SDN and optimizing the uptime of devices to building self-sustained networks that harvest energy from available sources.
\section{Use Cases of Maritime Communication Networks}
\label{IoT section}
This section presents some emerging use cases of maritime networks, such as the IoS, maritime IoT, and IoUT.
\begin{figure}[h]
    \centering
    \includegraphics[width=1\linewidth] {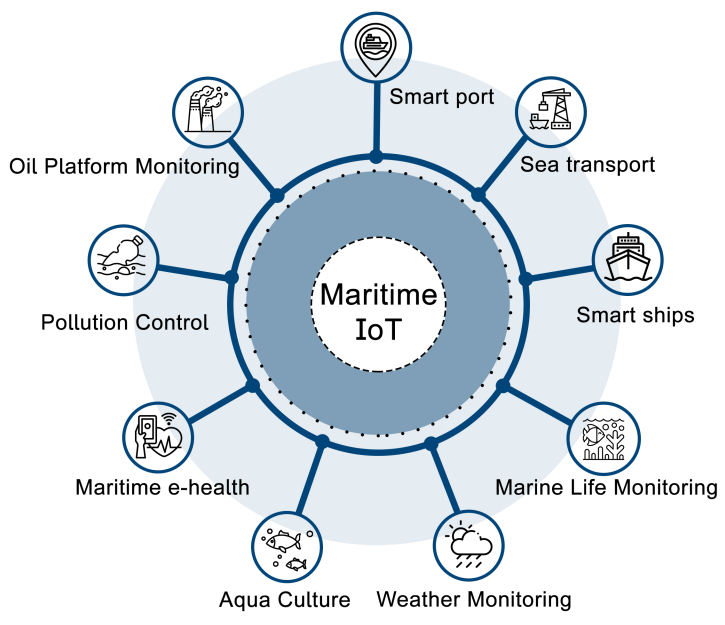}
    \caption {IoT categories and uses in maritime communications.}
    \label{Fig:IoTuses}
\end{figure}
\subsection{Internet-of-Ships Paradigm}
\label{Sec:IoSParadigm}

By anticipating IoT networks' economic and social advantages, the autonomous control of marine services can also bring new services. In the case of a maritime network, the nodes participating in developing the IoT setup are the devices of the network, such as ships and buoys, leading to the IoS paradigm \cite{martelli2020internet}. The IoS enables node computation coordination through some high-level virtualization of the core network where machine learning (ML) and artificial intelligence (AI) approaches are used to perform computational jobs linked to forecasting analysis.

According to \cite{liu2016internet}, the concept of IoS in shipbuilding might significantly influence ship construction and operation, with a wide range of future uses.
A more comprehensive study related to IoS is performed in \cite{aslam2020internet}, which examines three significant IoS subject areas: intelligent vessels, smart ports, and transportation. Reference \cite{aslam2020internet} also includes a discussion of the design, cores, and qualities that enable IoS unique to set it apart from other conventional IoT-based solutions. The IoS paradigm employs the AIS to establish marine operational capabilities and analyze maritime mobility. Nevertheless, effectively collecting salient data from the AIS dataset/database is usually a time-consuming and challenging task for maritime authorities and analysts. Hence, the authors of  \cite{he2017internet} proposed a new method to extract essential data from the AIS dataset by using a rule-based reasoning technique. Moreover, they also performed experiments at Yangtze River (China) to prove their findings, indicating that the suggested approach can extract data with great precision.

\subsection{Maritime IoT}
\label{Sec:IoTMaritime}
\begin{table*}
\caption{\label{Tab:IoTbasedProjects}IoT-based projects for maritime environments monitoring.}
\centering
\begin{tabular}{|p{1cm}|p{1cm}|p{3cm}|p{3.4cm}|p{7cm}|} 
\hline
Ref. & Year& Location & Communication Technology & Specifications \\
\hline
\cite{seders2007lakenet}&2007& Notre Dame (US)   &  Unlicensed UHF (433 MHz)&A sensing network, known as LakeNet, composed of 8 sensor pods, was deployed in a lake to monitor the water parameters.\\
\hline
\cite{regan2009demonstration}&2009&River Lee, Cork (Ireland)& ZigBee &A multi-sensor system was deployed for real-time water quality monitoring (by providing readings of water temperature, pH, oxygen level, and turbidity). \\
\hline
\cite{jin2010novel}&2010& Designed in China but not deployed    &  ZigBee/GPRS &A multi-sensor architecture to measure water parameters was proposed.  \\
\hline
{\cite{perez2011system}}&2011&  Mar Menor Golf (Spain)    &  ZigBee/GPRS &A WSN composed of four buoys deployed to record data on water temperature, pressure, and salinity, among other parameters.  \\
\hline
\cite{adamo2014smart}&2014&Adriatic Sea (Italy)   &  GPRS &  Deployment of an acoustic sensor network for water quality monitoring by measuring chlorophyll concentration together with other water physical parameters (temperature, turbidity, and salinity). \\
\hline
{\cite{ferreira2017autonomous}}&2017&  Atlantic Ocean (Portugal)    &  IEEE 802.11a/b/g/n and GPRS/UMTS/LTE &  Sea trials on connecting autonomous surface vehicles and autonomous underwater vehicles using a helikite-based BLUECOM+ network. \\
\hline
{\cite{al2018building}}&2018&North Sea (UK)& VHF & An IoT network architecture consisting of collecting data from marine sensory installed on ships is proposed. The gathered data is forwarded to onshore base stations. \\
\hline
{\cite{mourya2018ocean}}&2018&(Not Deployed)& Acoustic &A framework for oceanic spatio-temporal monitoring using acoustic sensor networks collecting underwater physical parameters (temperature, salinity, oxygen level, etc.) is proposed. \\
\hline
{\cite{morozs2018robust}}&2018& Fort William (UK)&Acoustic/TDA-MAC & Low-cost underwater acoustic sensor network deployment incorporating a TDA-MAC protocol for data collection. Several modifications were conducted to the TDA-MAC protocol to make it more robust in real-world deployments.\\ 
\hline
\end{tabular}
\end{table*}
Over the last few years, numerous research groups have conducted maritime IoT projects. A wireless sensor network (WSN) for marine environment monitoring based on the Zigbee technology was designed and deployed in Mar Menor coastal lagoon (Spain) \cite{perez2011system,PerezJOE17}. The network comprises four sensor nodes, each consisting of a solar-powered buoy equipped with air temperature and pressure sensors. The collected information is transmitted to a 30-km far base station \cite{PerezJOE17}.\newline
\indent Within the BLUECOM+ project (previously presented in \ref{StdWirelessAccessNet}), authors of \cite{ferreira2017autonomous} reported a series of oceanic life monitoring trials using autonomous robotic systems and sensors fixed on a drifter buoy. The data gathered during the tests that were conducted in the Portuguese coast-Atlantic Ocean was transmitted using the Helikite-based network for several kilometers. \newline
\indent Collecting information from IoT maritime sensor nodes can also be done through UAVs \cite{HuICCT18}. However, this method can be constrained by the UAVs' battery lifetime, mainly when the collection head node gathering the data from the sensor nodes is mobile. Authors of \cite{HuICCT18} proposed a routing maintenance approach for mobile sensor networks in maritime environments based on a ring broadcast mechanism consisting of finding the optimal path from sensor nodes to the collection head nodes and from the collection node to the nearest UAV.
\newline
In addition to oceanic data collection and observation, installing IoT inside shipping containers can enable monitoring shipments sensitive to temperature or humidity, for example. Authors of \cite{salah2020iot} designed and implemented a smart container prototype for shipping sensitive items with remote monitoring and location tracking capabilities. \newline
\indent Nonetheless, IoT-based solutions have a broad range of applications in marine environments, as seen in Fig. \ref{Fig:IoTuses}. For instance, consider a smart port, which enables authorities to provide their customers with more reliable information and innovative services \cite{yang2018internet}. Also, maritime IoT applications can serve in other situations like weather prediction, pollution control, and oil platform monitoring \cite{adiono2021internet, sai2020oil}. A summary of recent IoT projects and system deployments for monitoring maritime environments is given in Table \ref{Tab:IoTbasedProjects}. 

\subsection{Internet of Underwater Things}
\label{Sec:ShipUnderwaterIoT}
\begin{figure}[h]
    \centering
    \includegraphics[width=1\linewidth] {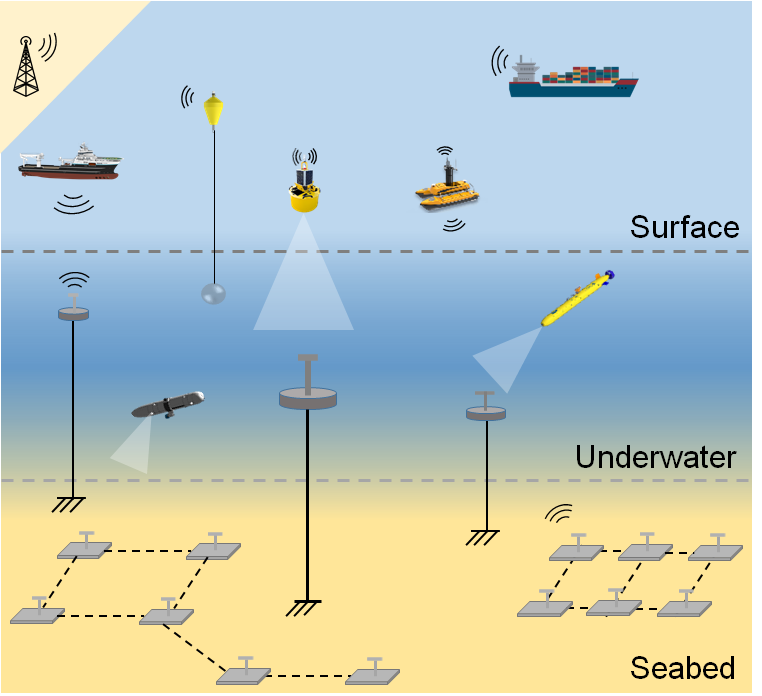}
    \caption {oUT network architecture.}
    \label{Fig:IoUTIllustation}
\end{figure}

Connecting objects underwater is essential for oil exploration, monitoring aquatic environments, and disaster prevention, among several other industrial and scientific applications. Ships, buoys, and autonomous surface vehicles (ASVs) can act as data collection stations, known as sinks, and gather data from underwater sensor networks (UWSNs) to transfer it to a control center via radio waves, as seen in Fig.~\ref{Fig:IoUTIllustation}.
UWSNs are traditionally based on acoustic communication to connect the different objects underwater \cite{akyildiz2004challenges, sendra2015underwater}. Other IoUT technologies involve using optical wireless communication \cite{jamali2016performance, zeng2016survey}, magnetic induction communication \cite{domingo2012magnetic, akyildiz2015realizing, Khalil2021}.
There are also hybrid technologies incorporating more than one type of communication for IoUT communication \cite{Celik2022, saeed2017energy}. Each of the IoUT communication technologies has its pros and cons. For instance, acoustic communication can transmit signals over long distances but is constrained by the limited bandwidth and the lack of stealth. The lack of reliability of acoustic IoUT is also a major challenge \cite{kao2017comprehensive}. The use of the optical band for IoUT networks benefits from the broad bandwidth, particularly around the blue-green (400-500 nm) region of the visible light spectrum.However, underwater optical communication can be strongly affected by turbulence from random variations in the water refractive index caused by temperature and salinity fluctuations. The transmission stability of magnetic induction communication is better than optical and acoustic communications, particularly in the air-water interface, as it is not affected by the index of refraction change. A comparison between the various ship-to-underwater IoT communication technologies' pros and cons is given in Table~\ref{Tab:TechnologiesComparison}. We note that RF communication is not included in the comparison (of Table~\ref{Tab:TechnologiesComparison}) due to the high attenuation of RF waves in seawater \cite{LacovaraMTS08} making the use of RF not possible in IoUT applications \cite{qiu2019underwater}. We also note that it is possible to have wired IoUTs connected with cables or optical fibers \cite{JahanbakhtCOMST21}.\newline
Regardless of the communication technology, information collection from IoUT devices is challenging, and a large portion of the collected information is not useful due to the sparse feature of the ocean \cite{XiangwangIEEENet21}. Optimizing the trajectories of AUVs used to collect data generated by IoUT to upload it to surface information collection stations is becoming a trendy topic \cite{FangIEEEIoT21}.\newline
Providing continuous powering to \textit{non-wired} IoUTs (as well as floating maritime IoT) or changing their batteries can be challenging. For such a reason, many devices were designed to harvest energy from sources at the seas, solar (light) using photovoltaic cells \cite{FilhoIEEECommag20} or wave using triboelectric generators \cite{BaiNanoEng19,ChandrasekharNanoEng20}. It is also possible to power IoUTs and maritime IoT by harvesting energy from wind or underwater currents. 
\begin{table*}
\caption{\label{Tab:TechnologiesComparison}Pros and cons of IoUT communication technologies}
\centering
\begin{tabular}{|p{2.4cm}|p{6.8cm}|p{7cm}|}
\hline
Technology & Pros & Cons \\
\hline
Acoustic&- Long reach of several hundred kilometers\newline - Has relaxed PAT requirements&- Limited bandwidth\newline - Subject to long delays due to the low sound velocity in the water\newline- Subject to multipath effect from reflection and refraction\newline - Noisy and can affect marine life and sea mammals\newline - High terminal power consumption \\
\hline
Optical Wireless&- Extended bandwidth with low absorption in the Blue-Green region of the visible light spectrum\newline- Stealth and without impact on marine life\newline- Small footprint and energy-efficient terminals&- Limited reach of a few tens of meters\newline - Subject to underwater turbulence \newline - Strong alignment requirement\\
\hline
Magnetic Induction&- Immune to underwater turbulence\newline - Practical for air-to-water links&- Limited propagation distance\newline
- Low data rate\\
\hline
\end{tabular}
\end{table*}
\subsection{Lessons Learned}
The applications of IoS and maritime IoT are growing. Having connected devices to facilitate navigation and data collection is becoming crucial for the naval industry and scientific research. Collecting information from floating IoT devices requires further UAVs and ASVs' implications. IoUT technology is also receiving significant attention as it can offer many opportunities to discover the underwater world and perform crucial sensing operations. Although many IoUT devices do not generate new data at a high frequency, connecting IoUT is needed and can be accomplished through acoustic, magnetic induction, and OWC communication. Each of these technologies has advantages and limitations. A common challenge for maritime IoT and IoUTs is providing continuous powering or changing batteries; therefore, these devices need to be designed to harvest energy from the available energy sources in the sea.   
\section{Challenges and Future Research Directions}
\label{Sec:Challenges}
Data acquired in the maritime sector could be inadequate, imprecise, or untrustworthy at specific periods or places due to the continuous mobility of ships and limited connectivity coverage in the sea. Naval vessels, for instance, are sometimes not linked to offer real-time data, and data may be dropped or interrupted due to a bad connection. These issues often obstruct the marine industry's ability to make timely and informed decisions. The shipping sector, for instance, needs to adopt new communication and data collecting technologies to cope with these critical challenges \cite{aslam2020internet}. Currently, most communications between ships and ships-to-shores when far from land are carried out through satellite communication \cite{hu2010applications}. However, satellite connections are costly and cause significant communication delays due to long propagation distances. Nevertheless, owing to the growing nature of marine applications, maritime network infrastructure is necessary to enable worldwide connectivity for ships, primarily across open seas and in the most distant parts of the world. Due to the importance of maritime networks, technological advancements have been carried out recently, which we presented in the previous sections; however, there are still many open research areas. 
Compared to terrestrial networks, maritime communication limited to a few Mbps in the best scenarios is still lagging behind in fulfilling the 5G requirements. 
In the following, we examine some open problems associated with maritime communication, and we present numerous research directions to connect vessels and improve on-board connectivity. Enabling additional features with communication is discussed.
\subsection{Safety and Security}
Maritime transportation is a safety-critical activity, but there is no standardized strategy in terms of cyber security in place \cite{jensen2015challenges}. It is also challenging to set cybersecurity standards in a short time in the maritime industry due to the lack of technical expertise in maritime IT departments and because a single shipping line could involve multiple entities in different locations. Cybersecurity attacks on shipping lines might result in severe outcomes, including maritime accidents and paralyzing supply chains. With the emergence of autonomous vessels, the impact of cyber attacks could lead to the worst consequences \cite{SilverajanAsiaJCIS}. For all these reasons, safety-critical network standards must be incorporated to improve the security of maritime networks \cite{wang2015big}.
\subsection{Bringing Broadband Cellular Connectivity to Deep Sea}
Regular mobile phones cannot be connected to terrestrial cellular broadband networks when far offshore (the maximum reach is between 20 and 25 km from shore in many countries). Although many projects are aimed to provide offshore broadband internet, they are limited to a few countries on the Atlantic \cite{MareFi14,BlueComPlus}, and many of such projects did not make it to commercialization. User mobile devices operating on terrestrial networks cannot directly connect to satellites. To connect to a non-terrestrial network (NTN), proprietary user equipment (UE) or VSAT is needed.
Connecting UEs, such as 5G ones, to an NTN  can be possible, but after coping with a wide range of challenges \cite{SatelliteUE}, such as the requirement of low latency. Connecting to a GEO satellite with a fixed coverage leads to at least 240 ms latency due to the significant round-trip propagation distance ($\sim$78.000 km). For this reason, relying on emerging LEO satellite constellations (such as Telesat Lightspeed and SpaceX Starlink) orbiting a few hundred kilometers from Earth can satisfy the low latency requirement.\newline 
In contrast to GEO satellites, LEO satellites have substantial coverage variations in time and space (\cite{SatelliteUE}, and references therein). Each ground terminal needs to be handed to another satellite every few minutes. Also, because LEO satellites travel at higher speeds than vessels, a significant Doppler Effect is created \cite{AliTCOM98}. With the current user, mobile terminal technology will still require having a satellite modem to access the internet. By involving HAPS acting as relays, it might be possible for UE terminals to connect to LEO satellite-provided broadband networks.\newline
\indent The widespread use of UAVs for wireless communication can also contribute to bridging the 5G divide between those sailing on-board and users on land. A potential idea would be to use agile UAVs flying in proximity of vessels to enable on-demand maritime coverage in fixed sea lanes as an addition to satellite and on-shore base stations \cite{LiIEEEWC20}. In such a hybrid architecture, a UAV can connect to terrestrial base stations in coastal areas and relies on satellites for backhauling when far offshore. One challenge of this approach is that UAVs require continuous powering. Given the sparsity of vessels in shipping lanes, the scheduling of the UAVs can be optimized to be deployed on user demand \cite{LiIEEEWC20}. For large vessels and cruise ships, potential alternatives would be the use of tethered UAVs and helikites.  \newline
\subsection{On-board VLC Communication}
VLC is maturing rapidly and can potentially be part of the future sixth-generation (6G) technology and beyond. VLC is an unlicensed technology that can co-exist with the lighting infrastructure. The coverage of light-emitting diodes (LEDs) used as VLC sources is restricted to the illuminated users making this technology secure from eavesdroppers in neighboring rooms. VLC is also immune to electromagnetic interference with RF terminals. VLC or data transfer through illumination can find potential use for on-board maritime communication. Given the progress in underwater optical wireless communication (UWOC) that operates with wavelengths in the visible spectrum \cite{NasirUWOCSurvey19}, using VLC in maritime communication can allow vessel communication with divers and remotely operated underwater vehicles (particularly when lasers are used as light sources). The main issue in this scenario will be fulfilling the pointing, acquisition, and tracking (PAT) requirements, particularly the potential for random movements in the air-water interface. Recently developed solutions based on scintillating fibers to relieve the PAT requirements for UWOC can be implemented to fulfill air-water (vessel-underwater) convergence \cite{SaitOPEX21}. Receivers with an enlarged field of view can be equally useful in similar situations where alignment is an issue \cite{AlkhazragiOL21}. In addition to scintillating fibers, extending the FoV of photonic receivers can be accomplished using fused optical fiber tapers \cite{AlkhazragiOL21} and luminescent solar concentrators (\cite{AlkhazragiSPIE22}, and references therein). VLC on-board can enable simultaneous lightwave information and power transfer (SLIPT) \cite{DiamantoulakisIEEETGCN18}. The illuminated user (i.e., under the LED coverage) can be charged by the information-carrying light signals, extending the working time of battery-based on-board IoTs. The SLIPT capability can also enable the charging of IoT devices used to collect climate variables \cite{JoseOPEX22}.
\subsection{A Room for THz Communication?}
The use of the THz band (from 0.1 THz to 10 THz) is envisioned as one of the enabling technology of the upcoming 6G era \cite{SaadIEEENet20}. THz signals are strongly absorbed by water vapor in the atmosphere making the use of signals in this band unsuitable for relatively long-range ship-to-ship and shore-to-ship applications. However, THz can benefit on-board applications requiring high data rates compared to what could be provided by microwave and technologies operating at lower RF frequency ranges. More importantly, THz can open the room for sensing on top of the communication as THz can be used for metal and gas sensing applications \cite{THzApplicationsCommag19, THzCommag20}. Potential THz maritime sensing use cases may involve sensing chemical leaks of biological materials on-board or around the vessel or an offshore oil platform. In addition to sensing, using the THz band can enable high-accuracy localization capabilities using small footprint antennas \cite{THzLocalization18}. This opens the opportunity to \textit{piggyback seamlessly} the localization feature into the on-board THz communication network. 
The use of the emerging \textit{holographic} intelligent reflecting surfaces (IRS) can facilitate the integration of these applications on top of the communication \cite{HuangIEEEWC}. IRS can also help improve THz's NLoS penetration, dominated by a strong LoS component \cite{ChaccourCOMST22}. 
\subsection{Harnessing the Power of Machine Learning for Maritime Communication}
There has been tremendous progress in using machine learning and related algorithms in RF and optical wireless communication. For instance, deep learning algorithms can be helpful when the channel is partially (or completely) unknown or difficult to model analytically either in RF \cite{YeLWC18} or FSO \cite{DarweshAccess20}. Deep learning algorithms have also been widely used to optimize coding and modulation in an end-to-end manner \cite{FelixSPAWC18}. Harnessing the power of ML can be beneficial for maritime communication using RF and optical waves.
ML algorithms can be practical for channel modeling and estimation under marine conditions. For instance, using a feedforward neural network, an ML supervised learning approach, can improve maritime wireless channel prediction under the atmospheric ducting phenomena as demonstrated \cite{ZhangICC21}. In \cite{LionisPhotonics21}, ML algorithms were used to predict the optical power of a maritime FSO link accurately. ML can be utilized to optimize coding and modulation in maritime communication and speed up complex operations.
One other use case of ML in maritime networks is enabling switching between RF and FSO links in hybrid radio/optical systems. Some ML algorithms, such as generative adversarial networks (GANs), can also help augment the data obtained from channel measurements needed for data-driven modeling. We note that GANs are machine learning-based frameworks that learn to generate new data with the same statistics as their training sets \cite{GANCommag,BriantcevOE22}. The benefits of these algorithms may cover simulation scenarios beyond what can be obtained with theoretical modeling and experimental measurements, particularly if we want to utilize signals in the optical band and potentially in the THz band. Leveraging data-driven learning can further enable joint sensing and communication as advocated in terrestrial networks \cite{AlkhateebArxiv22}. 
Reinforcement learning (RL), an ML paradigm widely applied for decision-making problems, can also improve the performance of UAV-based maritime communications. RL can help solve many issues of UAV networks, such as autonomous path planning and optimizing power consumption in a marine environment, as previously discussed in \cite{LahmeriIEEEOJCS21} for inland applications. Federated learning (FL) is another ML paradigm that can be useful for maritime IoT. For instance, each node or maritime IoT device can develop its own model in a decentralized manner to build a central model without sharing learning data with other nodes, which is helpful in the case of security risks and the absence of reliable links. Plenty of other opportunities can be offered through ML in maritime IoT \cite{YangIEEENet20}, which require further efforts. 
\subsection{Inter-Medium Communications}
In maritime communication, links across two different media (from water to air or vice versa) are equally important to cover in this survey. Few recent studies investigate inter-medium communications using relays or direct links in maritime networks. In the case of relays, the relay gives access to the other medium in decode and forward (DF) fashion while changing the communication technology. A demonstration from the literature reports using a photoacoustic device, which receives the information from the air by laser beam, and forwards the signal as acoustic wireless to the underwater receiver \cite{ji2021photoacoustic}. In \cite{rhodes2011underwater}, a design of a water surface station acting as a relay between air and underwater media was proposed. RF transceivers were mounted on the top of a buoy to receive RF signals and transmit them to the underwater medium using electrically insulated magnetic coupled antennas mounted on the bottom of the buoy \cite{rhodes2011underwater}. DF relaying can be beneficial for space-underwater communication. For instance, satellite signals in the microwave band cannot penetrate the water. Therefore they should be decoded at a floating station or a vessel and then forwarded to the submerged destination using another technology rather than microwave signals.\newline
 Inter-medium communication can be equally established using direct links. In the case of direct optical links, the diffused light in low turbid seawater can be detected by a photo-detector from underwater as demonstrated in \cite{sun2019realization}. The same method also can be conducted to establish underwater-to-air communication \cite{chen2021underwater}. Another way of establishing direct links is using vibration detection, where acoustic waves from the underwater environment vibrate the seawater up to the surface, which is detected via radar or laser Doppler vibrometer as demonstrated in \cite{tonolini2018networking}.  \newline
Inter-medium communications involve different fading effects (through the atmosphere and underwater) and crossing the air/water interface, which can severely affect the information propagating in direct links. This requires further investigation as the field of inter-medium communication with direct links is still largely unexplored for the various possible technologies, namely optical wireless, acoustic, and magnetic induction. \newline

\section{Conclusions}
 This article provides a state-of-the-art survey on maritime communications. We first provided an overview of maritime communication technologies based on radio bands and the optical spectrum. Different channel models for radio and optical wireless maritime links are studied. We also categorized the channel models depending on radio link communication scenarios and the weather conditions in free-space optics. We further covered different aspects of maritime networks, including modulation and coding schemes, radio resource management, coverage and capacity, and energy efficiency.
Moreover, we presented major use cases of IoT-related maritime networks. Compared to terrestrial communication, MCNs still lack high-speed links. Marine communication has been, most of the time, limited to the exchange of navigational information and critical data. Maritime communication for civil use can be subject to security bridges.
Bringing broadband connectivity to deep seas is another open challenge requiring further efforts. We finally discussed exciting research problems, including incorporating visible light and THz spectra in on-board applications. We stressed on leveraging the power of machine learning algorithms for maritime communication. Establishing reliable inter-medium communication is another area of focus. Boosting the role of machine learning in maritime communication and inter-medium communications. We believe this article provides valuable insights for maritime communications researchers in academia and industry and contributes to UN sustainable development goal 14 (``To conserve and sustainably use the oceans, seas and marine resources for sustainable development").
\label{Sec:Conclusion}
\section*{List of Acronyms}
\noindent 5G: 5th Generation\\
\noindent 6G: 6th Generation\\
\noindent AIS: Automatic Identification System\\
\noindent AO: Adaptive Optics\\
\noindent ASM: Application Specific Messages\\
\noindent BER: Bit Error Rate\\
\noindent B-LoS: Beyond Line-of-Sight\\
\noindent BS: Base Station\\
\noindent CH: Cluster Head\\
\noindent CN: Cluster Node\\
\noindent DPSK: Differential phase shift keying\\
\noindent DSC: Digital Selective Calling\\
\noindent DTN: Delay-Tolerant Networking\\
\noindent FDTD: Finite Difference Time Domain\\
\noindent FEM: Finite Element Method\\
\noindent FM: Frequency Modulation\\
\noindent FSO: Free Space Optics\\
\noindent GEO: Geostationary Earth Orbit\\
\noindent GAN: Generative Adversarial Network\\
\noindent GMSK: Gaussian Minimum Shift Keying\\
\noindent GNSS: Global Navigation Satellite System\\
\noindent GPRS: General Packet Radio Service\\
\noindent GPS: Global Positioning System\\
\noindent HAPS: High-Altitude Platform Station\\
\noindent HF: High Frequency\\
\noindent IM/DD: Intensity Modulation/Direct Detection\\
\noindent IMO: International Maritime Organization\\
\noindent IoS: Internet-of-Ships\\
\noindent IoT: Internet of Things\\
\noindent IoUT: Internet of Underwater Things\\
\noindent IRS: Intelligent Reflecting Surface\\
\noindent ITU: International Telecommunication Union\\
\noindent LED: Light Emitting Diode\\
\noindent LEO: Low Earth Orbit\\
\noindent LO: Local Oscillator\\
\noindent LTE: Long-Term Evolution\\
\noindent LoS: Line-of-Sight\\
\noindent MagicNet: Maritime Giant Cellular Network\\
\noindent MCN: Maritime Communication Network\\
\noindent MF: Medium Frequency\\
\noindent MIMO: Multiple-Input-Multiple-Output\\
\noindent MMR: Modulating Retro-Reflector\\
\noindent MMSI: Maritime Mobile Service Identity\\
\noindent mmWave: millimeter wave\\
\noindent NAVDAT: Navigation Data\\
\noindent NAVTEX: Navigational TEleX\\
\noindent NLoS: Non Line-of-Sight\\
\noindent OAM: Orbital Angular Momentum\\
\noindent OOK: On-Off Keying\\
\noindent PAM: Pulse Amplitude Modulation\\
\noindent PAT: Pointing, Acquisition, and Tracking\\
\noindent PE: Parabolic Equation\\
\noindent PPM: Pulse Position Modulation\\
\noindent QAM: Quadrature Amplitude Modulation\\
\noindent QoS: Quality of Service\\
\noindent QPSK: Quadrature Phase Shift Keying\\
\noindent RF: Radio Frequency\\
\noindent RRH: Remote Radio Head\\
\noindent RRM: Radio Resource Management\\
\noindent SDN: Software-Defined Networking\\
\noindent SISO: Single-Input-Single-Output\\
\noindent SLIPT: Simultaneous Lightwave Information and Power Transfer\\
\noindent SOTDMA: Self-Organized Time-Division Multiple Access\\ 
\noindent SNR: Signal-to-Noise Ratio\\
\noindent TDA-MAC: Transmit Delay Allocation MAC\\
\noindent TRITON: TRI-media Telematic Oceanographic Network\\ 
\noindent UAV: Unmanned Aerial Vehicle\\
\noindent UE: User Equipment\\
\noindent UHF: Ultra High Frequency\\
\noindent UMTS: Universal Mobile Telecommunications Systems\\
\noindent USV: Unmanned Surface Vehicle\\
\noindent UWOC: Underwater Wireless Optical Communication\\
\noindent VLC: Visible Light Communication\\
\noindent VHF: Very High Frequency\\
\noindent VSAT: Very Small Aperture Terminal\\
\noindent WiMAX: Worldwide Interoperability for Microwave Access \\
\noindent WLAN: Wireless Local Area Network\\
\noindent WSN: Wireless Sensor Network\\

\bibliographystyle{IEEEtran}
\bibliography{bibliography.bib}
\end{document}